\documentclass[
 reprint,
 amsmath,amssymb,
 aps,
]{revtex4-1}

\usepackage{graphicx}

\usepackage{subfigure}

\usepackage{url}

\begin{document}

\title{Fredkin and Toffoli gates implemented in Oregonator model \\ of Belousov-Zhabotinsky medium}
\author{Andrew Adamatzky}

 \email{andrew.adamatzky@uwe.ac.uk}
\affiliation{%
University of the West of England, Bristol, UK\\
}%

\date{\today}


\graphicspath{{figs/}}

\begin{abstract}
A thin-layer Belousov-Zhabotinsky (BZ) medium is a powerful computing device capable for implementing logical circuits, memory, image processors,  
robot controllers, and neuromorphic architectures. We design the reversible logical gates --- Fredkin gate and Toffoli gate --- in a BZ medium network of excitable  
channels with sub-excitable junctions.  Local control of the BZ medium excitability is an important feature of the gates' design.   A excitable thin-layer BZ medium 
responds to a localised perturbation with omnidirectional target or spiral excitation waves. A sub-excitable BZ medium responds to an asymmetric perturbation 
by producing travelling localised excitation wave-fragments similar to dissipative solitons. We employ interactions between excitation wave-fragments to 
perform computation. We interpret the wave-fragments as values of Boolean variables. A presence of a wave-fragment at a given site of a circuit represents  logical truth, absence of the wave-fragment ---  logical false. Fredkin gate consists of ten excitable channels intersecting at eleven junctions eight of which are sub-excitable. Toffoli gate consists of  six excitable channels intersecting at six junctions four of which are sub-excitable.  The designs of the gates are verified using numerical integration of 
two-variable Oregonator equations.
\end{abstract}


\maketitle

\section{Introduction}

A thin-layer Belousov-Zhabotinsky (BZ) medium~\cite{belousov1959periodic, zhabotinsky1964periodic} exhibits 
target waves, spiral waves and localised wave-fragments and their combinations. A number of theoretical and experimental laboratory prototypes of BZ computing devices 
have been produced. They are image processes and memory devices~\cite{kuhnert1986new, kuhnert1989image, kaminaga2006reaction},  logical gates implemented in geometrically constrained BZ medium~\cite{steinbock1996chemical, sielewiesiuk2001logical}, approximation of shortest path by excitation waves~\cite{steinbock1995navigating, rambidi2001chemical, adamatzky2002collision}, memory in BZ micro-emulsion \cite{kaminaga2006reaction}, information coding with frequency of oscillations~\cite{gorecki2014information}, onboard controllers for robots~\cite{adamatzky2004experimental, yokoi2004excitable, DBLP:journals/ijuc/Vazquez-OteroFDD14}, chemical diodes~\cite{DBLP:journals/ijuc/IgarashiG11}, neuromorphic architectures~\cite{ gorecki2006information, gorecki2009information, stovold2012simulating, gentili2012belousov, takigawa2011dendritic, stovold2012simulating,  gruenert2015understanding} and associative memory \cite{stovold2016reaction,stovold2017associative}, wave-based counters~\cite{gorecki2003chemical}, and  other information processors~\cite{DBLP:journals/ijuc/YoshikawaMIYIGG09, escuela2014symbol, gruenert2014understanding, gorecki2015chemical}. First steps have been already made towards prototyping arithmetical circuits with BZ: simulation and experimental laboratory realisation of gates~\cite{steinbock1996chemical, sielewiesiuk2001logical, adamatzky2004collision, adamatzky2007binary, toth2010simple, adamatzky2011towards},  clocks~\cite{de2009implementation} and evolving logical gates~\cite{toth2009experimental}. A one-bit half-adder, based on a ballistic interaction of growing patterns~\cite{adamatzky2010slime}, was implemented in a geometrically-constrained light-sensitive BZ medium~\cite{costello2011towards}.  Models of multi-bit binary adder, decoder and comparator in BZ are proposed  in~\cite{sun2013multi, zhang2012towards, suncrossover, digitalcomparator}. 
These architectures typically employ network of `conductive' channels made agar saturation with the reaction solution, crossover structures as T-shaped coincidence detectors~\cite{gorecka2003t} and chemical diodes~\cite{DBLP:journals/ijuc/IgarashiG11}. Excitation patterns in the  light-sensitive BZ medium~\cite{vavilin1968effect} can be controlled with illumination, therefore it is possible to make `conductive' just by varying illumination of the medium.  By controlling excitability~\cite{igarashi2006chemical}  in different loci of the medium we can achieve substantial results, e.g. implement analogs of  dendritic trees~\cite{takigawa2011dendritic}, polymorphic logical gates~\cite{adamatzky2011polymorphic},  integer square root circuits~\cite{stevens2012time}.

 \begin{table}[!tbp]
\caption{Truth table of Fredkin and Toffoli gates.}
\begin{center}
\begin{tabular}{p{1cm}p{1cm}p{1cm}||p{1cm}p{1cm}p{1cm}p{1cm}p{1cm}}
	&	Input	&		&		&		&	Output	&		&		\\ 
	&		&		&		&	Fredkin	&		&	Toffoli	&		\\  \hline
$z$	&	$x$	&	$y$	&	$c$	&	$a$	&	$b$	&	$a$	&	$b$	\\ 
0	&	0	&	0	&	0	&	0	&	0	&	0	&	0	\\ 
0	&	0	&	1	&	0	&	0	&	1	&	0	&	1	\\ 
0	&	1	&	0	&	0	&	1	&	0	&	1	&	0	\\ 
0	&	1	&	1	&	0	&	1	&	1	&	1	&	1	\\ 
1	&	0	&	0	&	1	&	0	&	0	&	0	&	0	\\ 
1	&	0	&	1	&	1	&	1	&	0	&	0	&	1	\\ 
1	&	1	&	0	&	1	&	0	&	1	&	1	&	1	\\ 
1	&	1	&	1	&	1	&	1	&	1	&	1	&	0	\\ 

\end{tabular}
\end{center}
\label{truthtable}
\end{table}

 In our previous paper on a BZ fusion gate and an adder made from fusion gates~\cite{adamatzky2015binary} we developed a hybrid approach to design of BZ-based logical circuits. We keep channels excitable, to remove the need of monitoring the whole medium, and junctions sub-excitable. In the channels BZ medium behaves as a `classical' excitable medium while allows for localised, soliton-like, wave-fragments to interact in the sub-excitable junctions.  In present paper we apply our approach to design and numerically simulate logically reversible Boolean gates: Fredkin~\cite{fredkin2002conservative} and Toffoli~\cite{toffoli1980reversible} gates.

We adopt the following symbolic notations. Boolean variables $x$ and $y$ take values `0', logical {\sc False}, and `1', logical {\sc True}; $xy$ is a conjunction (operation {\sc and}), $x+y$ is a disjunction ({\sc or}), $x\oplus y$ is exclusive disjunction (operation {\sc xor}), $\overline{x}$ is a negation of variable $x$ ({\sc not}).  In all designs presented  excitation waves are initiated at top ends of input channels and propagate down to output channels. The input channels are labelled by $x$, $y$, $z$ and output channels by   $a$, $b$, $c$.

The Fredkin and Toffoli gates have three inputs and three outputs each. Their truth tables are shown in Tab.~\ref{truthtable}. 
 Both gates transform input signals $x$ and $y$ controlled by signal $z$ 
to output signals $a$, $b$ and $c$. In both gates $c=z$, that is a control signal leaves the gate `unchanged' and $a$ and $b$ are calculated as follows: 
\begin{itemize}
\item Fredkin gate: $a = xz + y\overline{z}$ and $b=x\overline{z}+yz$
\item Toffoli gate: $a=x$ and $b=y\oplus(zx)$
\end{itemize} 
 Input variables $x$ and $y$ are recombined with the control signal.  In Fredkin gate an output is a disjunction of two conjunctions: 
 one input variable and control signal, another input variable and a negation of a control signal. In Toffoli gate one output is an identity of an input and second output is an 
 exclusive disjunction of one input variable with conjunction of another input variable with a control signal. The gates are important because they are key elements of low-power  computing circuits~\cite{de2011reversible, bennett1988notes} and they are amongst key components of quantum and optical computing circuits.
 These gates were implemented in theoretical and simulation design in optical~\cite{shamir1986optical, cuykendall1987control, milburn1989quantum, lloyd1993potentially, cerf1998optical, poustie2000demonstration, hardy2007optics, roy2010novel}, quantum~\cite{berman1994quantum, barenco1995elementary, smolin1996five, zheng2013implementation}, 
single electron~\cite{zardalidis2004design, wei2013compact},  and nano-mechanical systems~\cite{wenzler2013nanomechanical}; membrane P-systems~\cite{leporati2004simulating},  
DNA~\cite{seelig2006enzyme}, magnetic bubbles~\cite{chang1982magnetic}, enzymatic reactions~\cite{klein1999biomolecular, wood2004fredkin}, and slime mould~\cite{schumann2015conventional}. Experimental laboratory prototypes were produced with optical systems~\cite{kostinski2009experimental}, 
nuclear magnetic resonance~\cite{cory1998nuclear, fei2002realization, murali2002quantum}, and enzymatic reactions~\cite{moseley2014enzyme, orbach2014dnazyme, fratto2015reversible, fratto2015biomolecular, fratto2016controlled, fratto2016utilization, katz2017enzyme}.

The paper is structured as follows. The Oregonator model is described in Sect.~\ref{model}. 
Section~\ref{design} outlines design principles of the logical circuits.  Functioning of the gates is presented in full details in Sects.~\ref{fredkinsection} and \ref{toffolisection}. 
Physical reversibility of BZ circuits is discussed in Sect.~\ref{reversibility}.
Further developments are outlined in Sect.~\ref{discussion}.

\section{Oregonator model of sub-excitable medium}
\label{model}

\begin{figure}[!tbp]
\centering
\includegraphics[scale=0.35, angle=90]{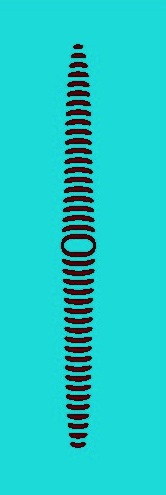}
\caption{ 
Time lapsed snapshots of two excitation wave-fragments in sub-excitable medium. 
The grid of  500$\times$500 node is excited with a rectangular domains 3$\times$20 nodes at the centre of a grid. 
This excitation produces two excitation wave-fragments. One wave-fragment propagates to the left another to the right.
The picture is the time lapsed snapshot of the medium recorded every $150^{th}$ step of numerical integration.
See videos at \protect\url{https://drive.google.com/open?id=0BzPSgPF_2eyUVTMxbzFuYTZvVzg}
}
\label{mediumtypes}
\end{figure}

We use two-variable Oregonator equations~\cite{field1974oscillations} adapted to a light-sensitive  Belousov-Zhabotinsky (BZ) reaction with applied illumination~\cite{beato2003pulse}. The Oregonator equations in chemistry bear the same importance as Hodgkin-Huxley and FitzHugh-Nagumo equations in neurophysiology, 
Brusselator in thermodynamics,  Meinhardt-Gierer in biology,  Lotka-Volterra  in ecology, and Fisher equation in genetics. The Oregonator equations are used to model a wide range of phenomena in BZ, e.g.  analysis of rotating waves~\cite{jahnke1989chemical}, chaos in flow BZ \cite{tomita1979chaos},   stochastic resonance in BZ \cite{amemiya1998two}, effect of macro mixing \cite{hsu1994effects}. The Oregonator equations is the simplest continuous model of the BZ medium yet showing
very good agreement with laboratory experiments.  Let us provide few examples.   A stable three-dimensional organising centre that periodically emits trigger excitation waves   found experimentally is reproduced in the Oregonator model \cite{azhand2014three}. Studies of the BZ system with a global negative feedback demonstrate that the Oregonator model shows the same bifurcation scenario of bulk oscillations and wave patterns emerging when the global feedback exceed a critical value as the bifurcation scenario observed in laboratory experiments \cite{vanag2000pattern}. There is a good match between lab experiments on modifying excitation wave patterns in BZ using external DC field and the Oregonator model of the same phenomena~\cite{vsevcikova1996dynamics}. The Oregonator model used  in~\cite{dockery1988dispersion} to evaluate the dispersion relation for periodic wave train solutions in BZ shows agrees with experimental results. Patterns produced  by  the Oregonator model of a three-dimensional scrolls waves are indistinguishable from patterns produced in the  laboratory experiments~\cite{winfree1989three}. Excitation spiral breakup demonstrated in the Oregonator model is verified in experiments~\cite{taboada1994spiral}. The Oregonator model can be finely tuned, e.g. adjusted for temperature 
dependence \cite{pullela2009temperature}, scaled~\cite{hastings2016oregonator}, modified for oxygen sensitivity~\cite{krug1990analysis}. Author with colleagues personally used the Oregonator model as a fast-prototyping tool and virtual testbed in designing BZ medium based computing devices which were implemented experimentally~\cite{adamatzky2007binary, de2009implementation, toth2009experimental, toth2010simple, adamatzky2011towards, stevens2012time}.  

The Oregonator equations are following: 
\begin{eqnarray}
  \frac{\partial u}{\partial t} & = & \frac{1}{\epsilon} (u - u^2 - (f v + \phi)\frac{u-q}{u+q}) + D_u \nabla^2 u \nonumber \\
  \frac{\partial v}{\partial t} & = & u - v 
\label{equ:oregonator}
\end{eqnarray}
The variables $u$ and $v$ represent local concentrations of an activator, or an excitatory component of BZ system, and an inhibitor, or a refractory component. Parameter $\epsilon$ sets up a ratio of time scale of variables $u$ and $v$, $q$ is a scaling parameter depending on rates of activation/propagation and inhibition, $f$ is a stoichiometric coefficient. Constant $\phi$ is a rate of inhibitor production.  In a light-sensitive BZ $\phi$ represents the rate of inhibitor production proportional to intensity of illumination.
We integrate the system using Euler method with five-node Laplace operator, time step $\Delta t=0.001$ and grid point spacing $\Delta x = 0.25$, $\epsilon=0.02$, $f=1.4$, $q=0.002$, $D_u=1.0$, $D_v=0.0$.   To generate excitation waves of wave-fragments we perturb the medium by a square solid domains of excitation, $10 \times 1$ sites (unless otherwise stated) in state $u=1.0$.  
 The parameter $\phi$ characterises excitability of the simulated medium.
The medium is excitable, it exhibits `classical' target waves when $\phi=0.05$. The medium is non-excitable when $\phi>0.09$. The medium is sub-excitable with propagating localizations, or wave-fragments, when  $\phi=0.0766$ (Fig.~\ref{mediumtypes}).  

We guide wave-fragment by constraining them to excitable channels and by colliding the wave-fragments with each other at the junctions of the channels~\cite{adamatzky2015binary}. The channels used in the designs presented have width 30 nodes. At most junctions channels 
intersect at a right angle to keep `contact size' between the channels minimal and thus to reduce chances of a single wave propagating along one channel to `spread' into another channel. 

There are two type of junctions: excitable and sub-excitable. In the excitable junction, as in the channels, $\phi=0.05$. In the sub-excitable junction, 
each node being at a distance less than 15 nodes from the centre of the junction has $\phi=0.0766$, on the design schemes all nodes of the junctions inside 
the circles has $\phi=0.0766$.

There are three scenarios of a wave crossing the junction. If the junction is excitable ($\phi=0.05$) the wave spreads in all channels.  
If the junction is sub-excitable ($\phi=0.0766$) and only one wave approaches the junction, this lonely wave is transformed to a
 wave-fragment which crosses the junction conserving its shape. If two wave enter the sub-excitable junction, they collide and fuse into a single wave-fragment.

Time lapse snapshots provided in the paper were recorded at every 150 time steps, we display sites with $u >0.04$. Videos supplementing figures were produced by saving a frame of the simulation every 10th step of numerical integration and assembling them in the video with play rate 30 fps.

\section{Design principles}
\label{design}

\begin{figure}[!tbp]
\centering
\subfigure[]{\includegraphics[scale=0.35]{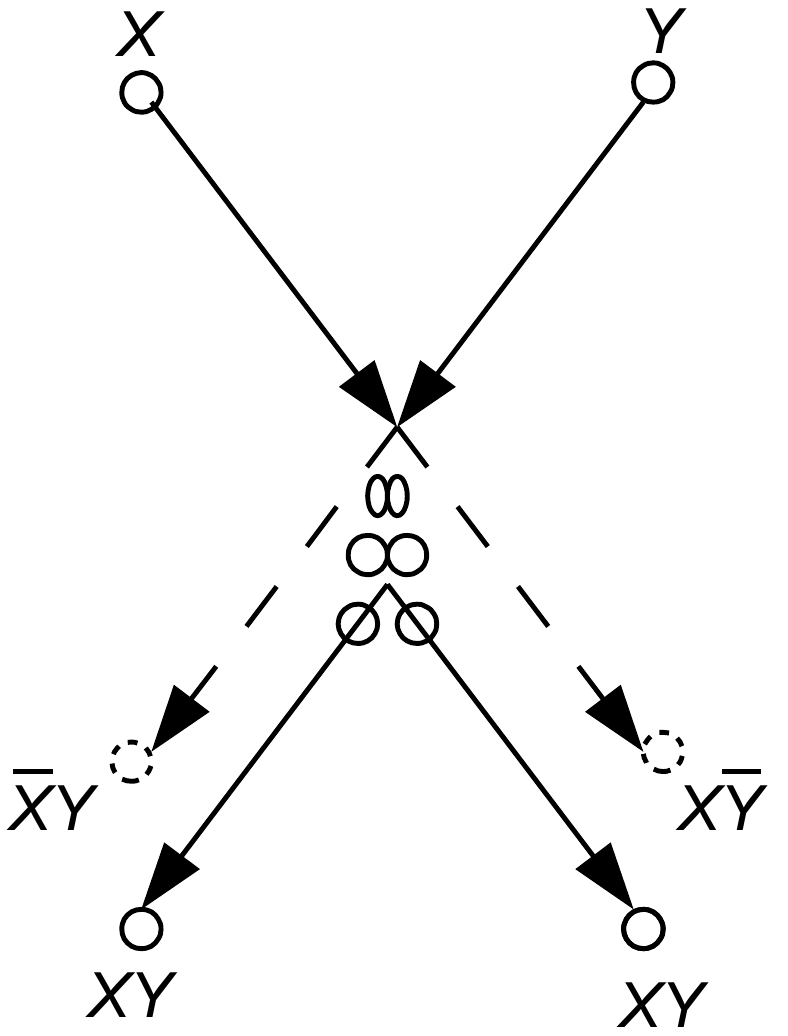}}
\subfigure[]{\includegraphics[scale=0.3]{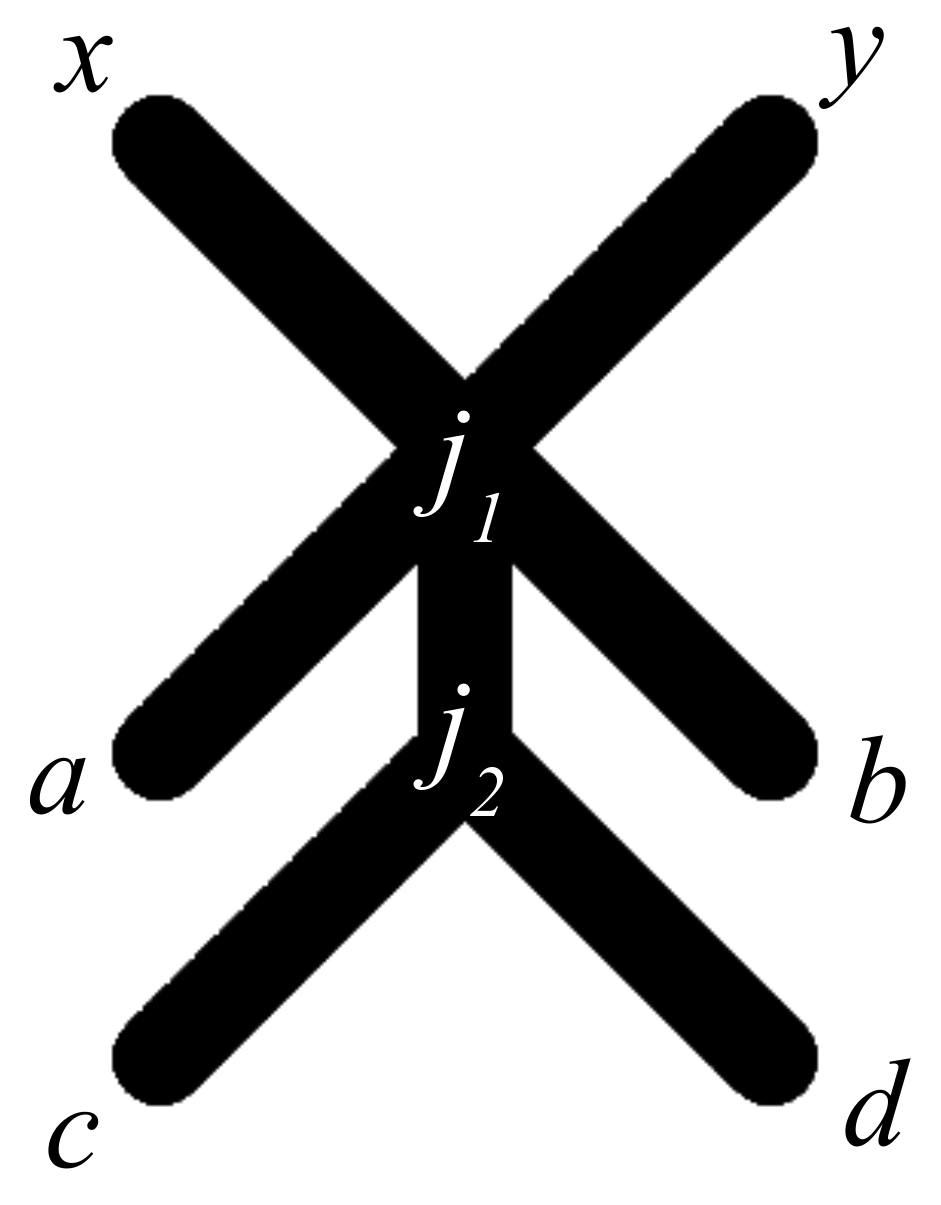}}
\subfigure[]{\includegraphics[scale=0.3]{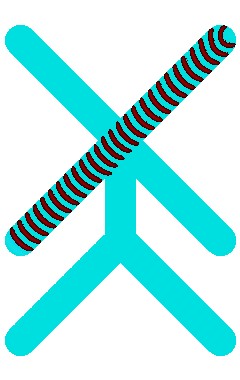}}
\subfigure[]{\includegraphics[scale=0.3]{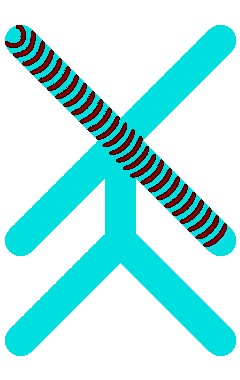}}
\subfigure[]{\includegraphics[scale=0.3]{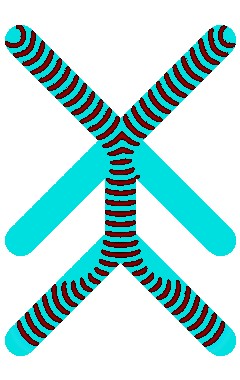}}
\caption{Time lapsed snapshots of wave-fragments propagating in
  simulated BZ medium implementing  Margolus gate. 
  Bounding box of the gate is 234 $\times$ 334 nodes.
(a)~Scheme of the gate.
(b)~$x=0$, $y=1$, 
(c)~$x=1$, $y=0$,
(d)~$x=1$, $y=1$.
}
\label{margolus}
\end{figure}

A  collision-based computation, inspired by Fredkin-Toffoli conservative logic~\cite{fredkin2002conservative}, employs mobile compact finite patterns which implement 
computation while interacting with each~\cite{adamatzky2002collision}. Information values (e.g. truth values of logical variables) are given by either absence or presence of the localisations or other parameters of the localisations. The localisations travel in space and perform computation when they collide with each other.  Almost any part of the medium space can be used for computation.  The localisations undergo transformations, they change velocities, form bound states, annihilate or fuse when they interact with other localisations. Information values of localisations are transformed in the collisions and thus a computation is implemented.  The concept of the collision-based logical gates with excitation wave-fragments is best illustrated using a gate based on collision between two soft balls, the Margolus gate~\cite{margolus2002universal}, shown in Fig.~\ref{margolus}a. Logical value $x=1$  is given by a ball presented in input trajectory marked $x$, and $x=0$ by absence of the ball in the input trajectory $x$; the same applies to $y=1$ and $y=0$. When two balls approaching the collision gate along paths $x$ and $y$ collide, they compress but then spring back and reflect. 
The balls come out along the paths marked $xy$. If only one ball approaches the gate, for inputs $x=1$ and $y=0$ or $x=0$ and $y=1$, the ball exits the gate via path $x\overline{y}$ (for input $x=1$ and $y=0$) or $\overline{x}y$ (for input $x=0$ and $y=1$).

The Margolus gate is implemented  by straightforward mapping of balls trajectories (Fig.~\ref{margolus}a) 
to a configuration of excitable channels  (Fig.~\ref{margolus}b). Junction $j_1$ is sub-excitable ($\phi=0.0766$). Junction $j_2$ is excitable ($\phi=0.05$).  Functioning of the gate is shown in Fig.~\ref{margolus}c--i and the representation of logical values on each segment of the gate in Tab~\ref{segmentvalues}a. If inputs are $x=0$ and $y=1$ the input channel $y$ excited  (Fig.~\ref{margolus}c). The wave-fragment 
propagates across junction $j_1$ into output $a$; the wave does not spread into $b$ because the junction $j_1$ is sub-excitable.  
If inputs are $x=1$ and $y=0$ the input $x$ is excited  (Fig.~\ref{margolus}d). The wave-fragment  propagates across junction $j_1$ into output $b$; 
the wave does not spread into $b$ because the junction $j_1$ is sub-excitable.   If inputs are $x=1$ and $y=1$ both input channels are excited (Fig.~\ref{margolus}e). The wave-fragments 
propagating along channels $x$ and $y$ collide at junction $j_1$: they fuse into a single wave-fragment which enters segment $j_1 j_2$. 
On reaching the excitable junction $j_2$ the wave-fragment splits into two wave-fragments. One wave propagates into segment $c$, another wave into segment $d$. 
 Excitation wave appears at output $a$ only if channel $x$ was not excited and channel $y$ was excited. Thus we have $a=\overline{x}y$.   Excitation wave appears at output $b$ only if channel $x$ was excited and channel $y$ was not excited. Thus we have 
$b=x\overline{y}$. Excitation wave propagating along segment $j_1 j_2$ imitates two soft balls propagating as a single body for some time after their collision (Fig.~\ref{margolus}a). Splitting of the wave-fragment into segments $c$ and $d$ (Fig.~\ref{margolus}e) imitates soft balls springing back and reflecting (Fig.~\ref{margolus}a):  $c=xy$ and $d=xy$.

\begin{table}[!tbp]
\caption{Representation of the logical values in segments of the BZ implementations of 
(a)~Margolus gate, (b)~Fredkin  gate, (c)~Toffoli gate.}
\centering
\subfigure[]{
\begin{tabular}{c|c}
Segment	&	Value 	\\ \hline
$xj_1$	&	$x$	\\
$yj_1$	&	$y$	\\
$j_1a$	&	$\overline{x}y$	\\
$j_1b$	&	$x\overline{y}$	\\
$j_1j_2$	&	$xy$	\\
$j_2c$	&	$xy$	\\
$j_2d$	&	$xy$	\\
\end{tabular}}
\subfigure[]{
\begin{tabular}{c|c}
Segment	&	Value 	\\ \hline
$xj_4$	&	$x$	\\
$zj_1$	&	$z$	\\
$j_1j_3$	&	$z$	\\
$j_1j_2$	&	$z$	\\
$j_2c$	&	$z$	\\
$j_3j_4$	&	$z$	\\
$j_3j_5$	&	$z$	\\
$yj_2$	&	$y$	\\
$j_2j_5$	&	$y$	\\
$j_4j_7$	&	$xz$	\\
$j_4j_6$	&	$x\overline{z}$	\\
$j_5j_6$	&	$y\overline{z}$	\\
$j_5j_9$	&	$yz$	\\
$j_6j_7$	&	$\overline{x}y\overline{z}$	\\
$j_6j_9$	&	$x\overline{y}\overline{z}$	\\
$j_7j_{10}$	&	$xz+\overline{x}yz$	\\
$j_9j_{11}$	&	$yz+x\overline{y}\overline{z}$	\\
$j_6j_8$	&	$xy\overline{z}$	\\
$j_8j_{10}$	&	$xy\overline{z}$	\\
$j_8j_{11}$	&	$xy\overline{z}$	\\
$j_9j_{11}$	&	$yz+x\overline{y}\overline{z}$	\\
$j_{10}a$	&	$xz+y\overline{z}$ \\ 
$j_{11}b$	&	$x\overline{z}+yz$ 
\end{tabular}
}
\subfigure[]{
\begin{tabular}{c|c}
Segment	&	Value 	\\ \hline
$xj_1$	&	$z$	\\
$j_1c$	&	$z$	\\
$j_1j_3$	&	$z$	\\
$yj_2$	&	$x$	\\
$j_2j_3$	&	$x$	\\
$j_2j_4$	&	$x$	\\
$j_4a$	&	$x$	\\
$j_3j_4$	&	$zx$	\\
$j_4j_5$	&	$zx$	\\
$j_5j_7$	&	$zx\overline{y}$	\\
$j_5j_6$	&	$\overline{zx}y$	\\
$j_6b$	&	$y\oplus (zx)$	\\
\end{tabular}
}
\label{segmentvalues}
\end{table}%

\section{Implementation of Fredkin gate}
\label{fredkinsection}

\begin{figure}[!tbp]
\centering
\includegraphics[scale=0.4]{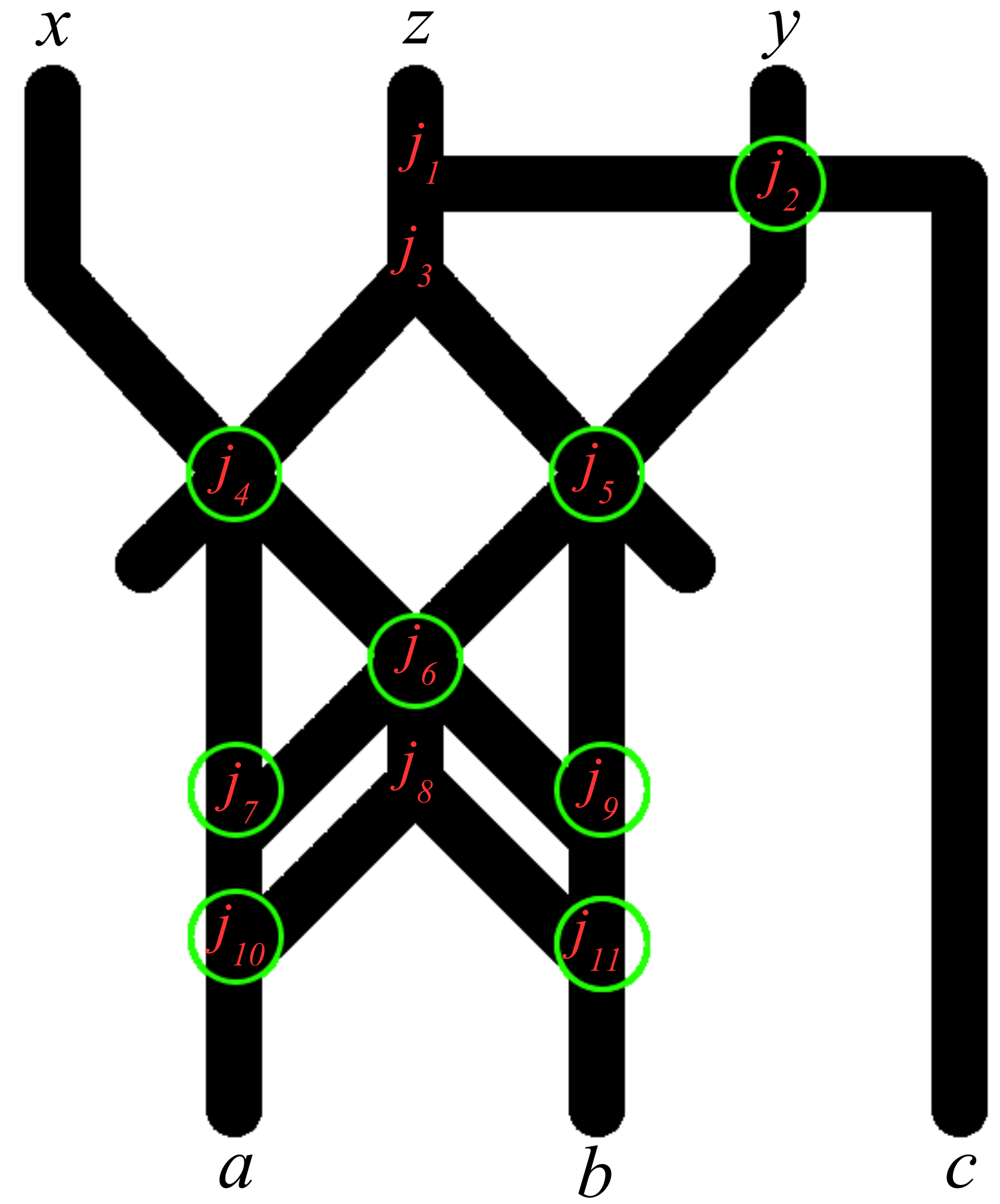}
\caption{Implementation of Fredkin gate. 
Black channels are excitable; white space is non-excitable. 
Junctions are labelled $j_1$ to $j_1$; sub-excitable junctions are encircled.
Bounding box of the taste is 532 $\times$ 532 nodes.}
\label{fredkinscheme}
\end{figure}

\begin{figure*}[!tbp]
\centering
\subfigure[]{\includegraphics[scale=0.2]{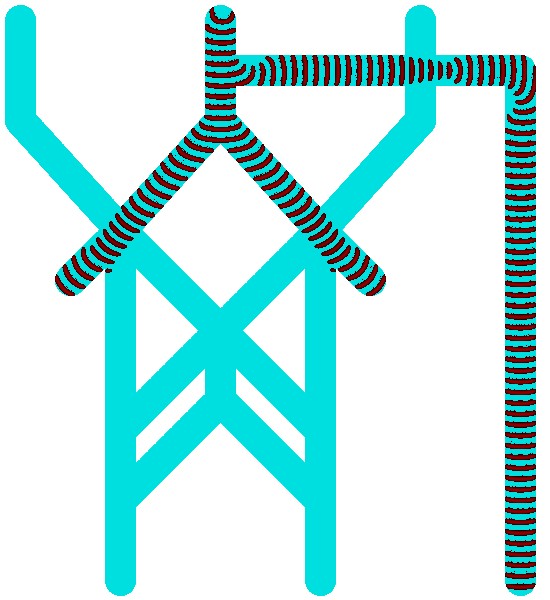}}
\subfigure[]{\includegraphics[scale=0.2]{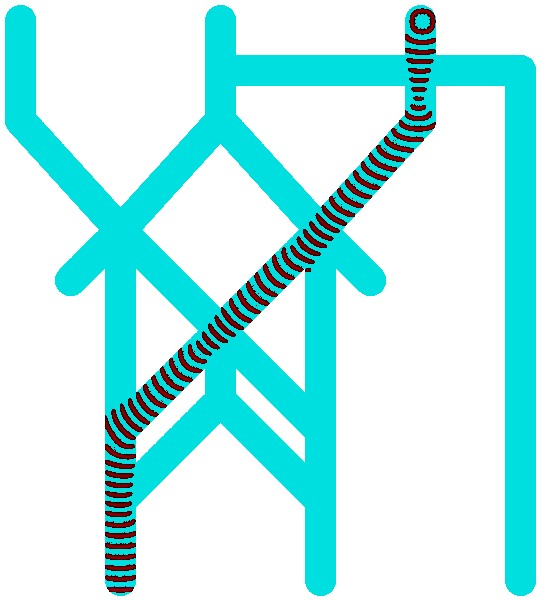}}
\subfigure[]{\includegraphics[scale=0.2]{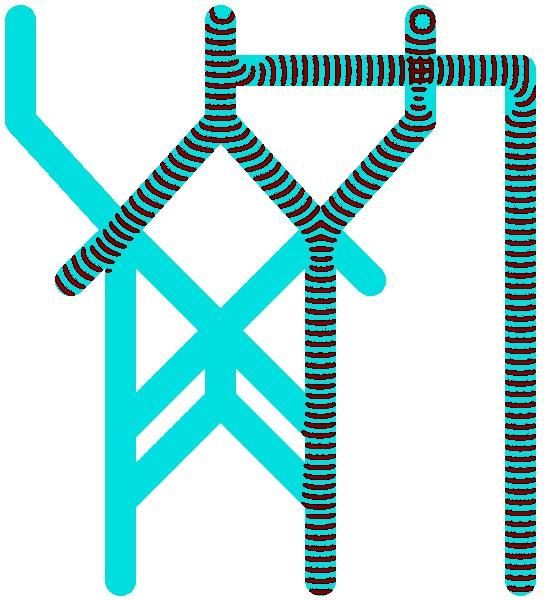}}
\subfigure[]{\includegraphics[scale=0.2]{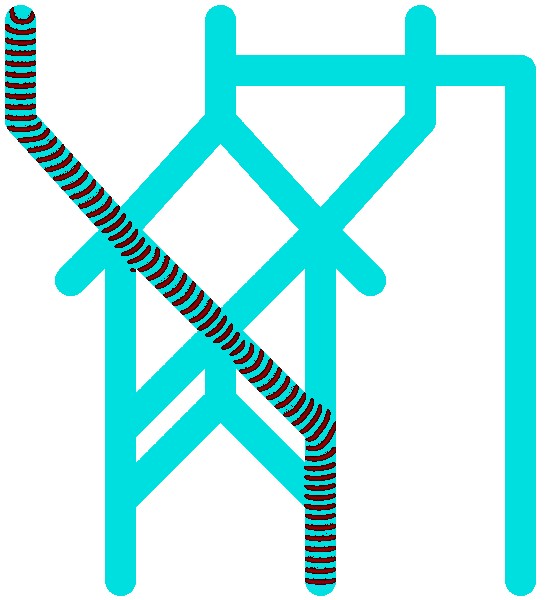}}
\subfigure[]{\includegraphics[scale=0.2]{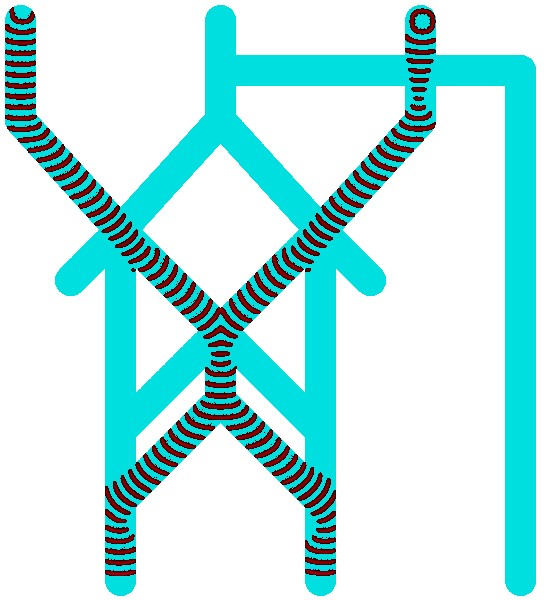}}
\subfigure[]{\includegraphics[scale=0.2]{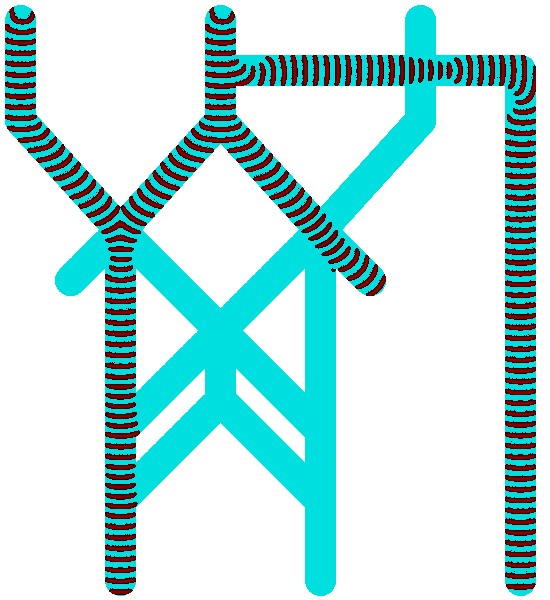}}
\subfigure[]{\includegraphics[scale=0.2]{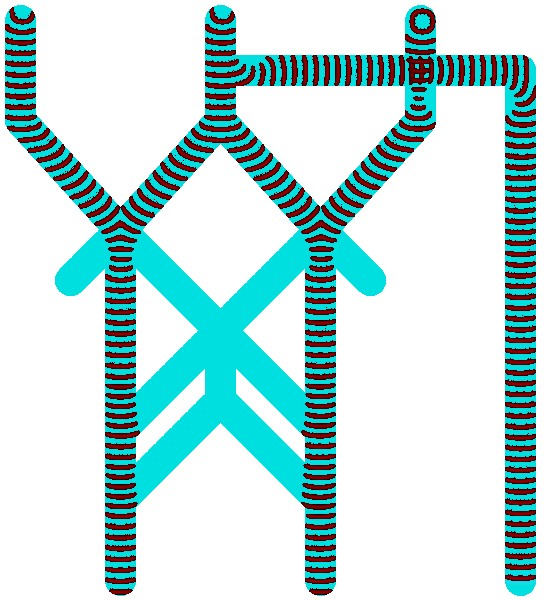}}
\caption{Time lapsed snapshots of wave-fragments propagating in
  simulated BZ medium implementing Fredkin gate. 
  (a)~$x=0, y=0, z=1$, 
  (b)~$x=0, y=1, z=0$,
  (c)~$x=0, y=1, z=1$,
  (d)~$x=1, y=0, z=0$,
  (e)~$x=1, y=0, z=1$,
  (f)~$x=1, y=1, z=0$,
  (g)~$x=1, y=1, z=1$.
See videos at \protect\url{https://drive.google.com/open?id=0BzPSgPF_2eyUSjM2aWZlU21fM3c}}
\label{fredkinsnapshots}
\end{figure*}

Design of excitable implementation of Fredkin gate consists of ten excitable channels intersecting at eleven junctions (Fig.~\ref{fredkinscheme}). All junctions but $j_1$, $j_3$, $j_8$
are sub-excitable. The exact domains of sub-excitable ares shown in circles in Fig.~\ref{fredkinscheme}. The gate implement transformation of the Boolean triple $(x,y,z)$ to $(a,b,c)$.
Inputs $x=1$, $y=1$ or $z=1$ are represented by excitations initiated at the entrances of the input channels; if an input variable is `0' the corresponding  input channel is not excited. 
Outputs $a$, $b$, $c$ are assumed to have value `1' when an excitation wave appears at the corresponding output.

Input $x=0$, $y=0$, $z=1$. Excitation wave is initiated at entrance of channel $z$ (Fig.~\ref{fredkinsnapshots}a).  The wave propagates along segment $zj_1$. 
Junction $j_1$ is excitable therefore the wave enters segments $j_1 j_3$ and $j_1 j_2$. The wave entered $j_1 j_2$ propagates across sub-excitable junction $j_2$ into segment $j_2 c$ without spreading into segment $j_2 j_5$; this wave reaches the output $c$. The wave propagating in segment $j_1 j_3$ splits into two waves at the excitable junction $j_3$. One wave propagates along $j_3 j_4$ and another along $j_3 j_5$.  Junctions $j_4$ and $j_5$ are sub-excitable therefore the waves run across the junctions and get extinguished in the cul-de-sacs.

Input $x=0$, $y=1$, $z=0$. Excitation wave is initiated in channel $y$. The wave crosses sub-excitable junction $j_2$ without spreading to neighbouring segments $j_1 j_2$ and $j_2 c$ but propagates only into $j_2 j_5$. Junctions $j_5$ and $j_6$ are sub-excitable. The excitation wave runs across these two junctions without spreading to neighbouring segments. 
When the wave reaches junction $j_{10}$ it collides with the wall of the segment $j_{10} a$, slightly reflect and proceeds towards output $a$ (Fig.~\ref{fredkinsnapshots}b).

Input $x=0$, $y=1$, $z=1$. Input channels $y$ and $z$ are excited. Excitation wave-fronts propagate in $z j_1$ and $y j_2$ (Fig.~\ref{fredkinsnapshots}c).  
The wave-front originated in $y$ travels across sub-excitable junction $j_2$ into segment $j_2 j_5$ without expanding into segments $j_1 j_2$ and $j_2 c$. The wave-front originated 
in $z$ spreads into $j_1 j_2$, $j_1 j_3$, $j_3 j_4$ (where the excitation dies) and $j_3 j_5$. The wave-fronts propagating along $j_3 j_5$ and $j_2 j_5$ collides 
at sub-excitable junction $j_5$. These two wave-fronts collide because distance from $z$ to $j_5$ equals to distance from $y$ to $j_5$. The wave-fronts merge on `impact' and fuse into a single excitation wave-front propagating into $j_5 j_9$. Junctions $j_9$ and $j_{10}$ are sub-excitable therefore the wave-front cross them without expanding, moving directly to 
output $b$. The wave-front propagating along $j_1 j_2$ crossed junction $j_2$ without expanding and reaches output $c$. Note that on the time-lapsed snapshots 
(Fig.~\ref{fredkinsnapshots}c) traces of wave-fronts travelling from $j_1$ to $c$ and form $y$ to $j_5$ intersect:  wave-fronts \emph{per se} do not collide because distance 
from $z$ to $j_2$ is large than distance from $y$ to $j_2$. See videos at \protect\url{https://drive.google.com/open?id=0BzPSgPF_2eyUSjM2aWZlU21fM3c}.

Input $x=1$, $y=0$, $z=0$. The excitation dynamics is analogous to the dynamics for inputs $x=0$, $y=0$, $z=1$. The wave-fragment initiated at $x$ propagates along along path $x$ to $j_4$ to $j_6$ to $j_9$ to $j_{11}$ to $b$ (Fig.~\ref{fredkinsnapshots}d).

Input $x=1$, $y=1$, $z=0$.  Excitation wave-front originated in $x$ collides with excitation wave-front originated in $y$ at junction $j_6$. The wave-front fuse into a single wave-front. This front 
travels along $j_6 j_8$. Junction $j_8$ is excitable therefore the wave-front spreads into segments $j_8 j_{10}$ and $j_8 j_{11}$ and to outputs $a$ and $b$ (Fig.~\ref{fredkinsnapshots}e).

Input $x=1$, $y=0$, $z=1$. Excitation initiated in $x$ propagates towards $j_4$ (Fig.~\ref{fredkinsnapshots}f). Excitation initiated in $z$ propagates along path $z j_1 j_2 c$ to output $c$ and also along path $j_1 j_3 j_4$ to junction $j_4$. Distance from $x$ to $j_4$ is the same as distance from $z$ to $j_4$. Therefore wave-fronts travelling along paths $x j_4$ and $z  j_4$ collide with each other. They fuse into a single excitation wave-front. This wave-front propagates to output $a$.

Input $x=1$, $y=1$, $z=1$. Excitations are initiated in all three inputs (Fig.~\ref{fredkinsnapshots}g). Wave-fronts originated in $x$ and $z$ collide with each other at junction $j_4$. They fuse into a wave-fragment travelling towards  output $a$.  Wave-fronts originated in $z$ and $y$ collide with each other at junction $j_5$. They fuse into a wave-fragment travelling towards  output $b$. The excitation wave-front from $z$ also travels to output $c$.   

Thus, we have $c=z$, $a=xz+y\overline{z}$, $b=x\overline{z}_yz$. Exact Boolean functions represented by a wave-fragment on each of the segments are shown in Tab.~\ref{segmentvalues}b.

In design Figs.~\ref{fredkinscheme}~and~\ref{fredkinsnapshots} the path from  $z$ to $c$ is longer than paths  from $x$, $y$ or $z$ to  $a$ and $b$ 
therefore a signal arrives at output $c$ later than signals arrived at outputs $a$ and $b$.  This can be amended by making output segments $a$ and $b$ longer by transforming them in zig-zag segments. We did not show this compensation on the scheme or video not to clutter the designs.

\section{Implementation of Toffoli gate}
\label{toffolisection}

\begin{figure}[!tbp]
\centering
\includegraphics[scale=0.35]{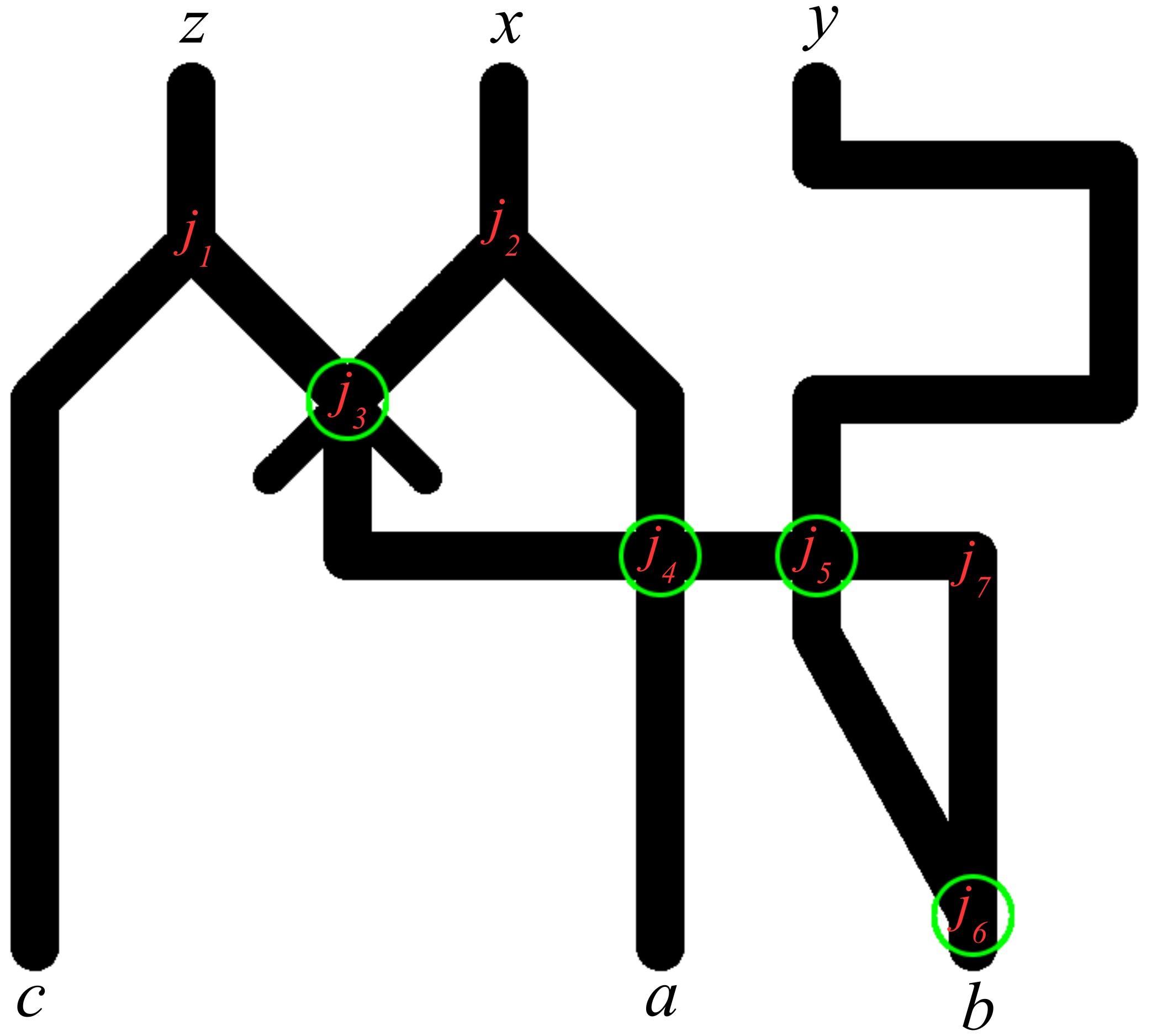}
\caption{Excitable medium implementation of Toffoli gate. Black channels are excitable and sub-excitable; white space is non-excitable. 
Junction are labelled $j_1$ to $j_7$; sub-excitable junctions are encircled.
Bounding box of the taste is 720 $\times$ 580 nodes.
}
\label{toffoli}
\end{figure}

\begin{figure*}[!tbp]
\centering
\subfigure[]{\includegraphics[scale=0.2]{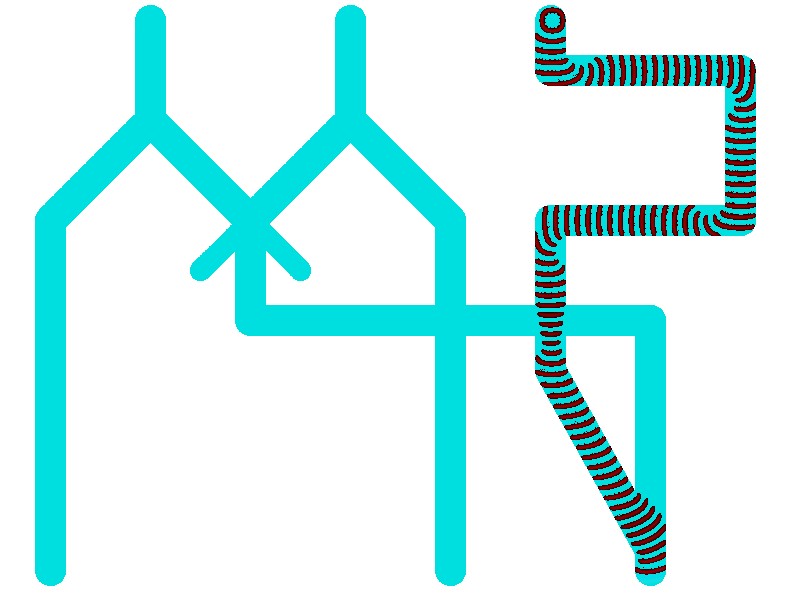}} 
\subfigure[]{\includegraphics[scale=0.2]{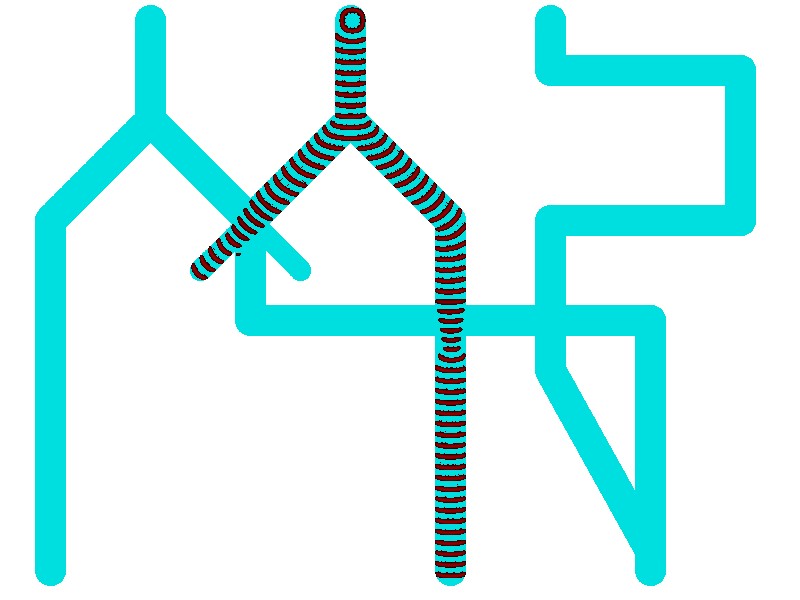}} 
\subfigure[]{\includegraphics[scale=0.2]{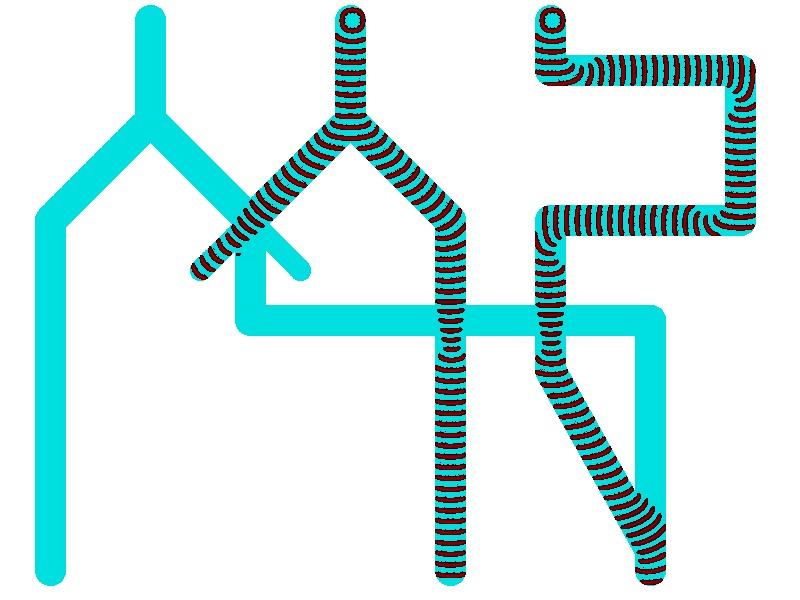}} 
\subfigure[]{\includegraphics[scale=0.2]{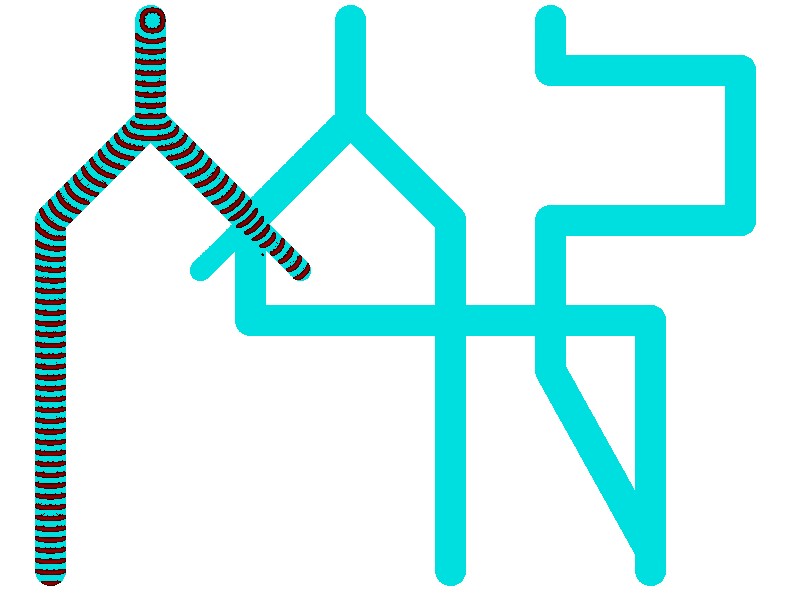}} 
\subfigure[]{\includegraphics[scale=0.2]{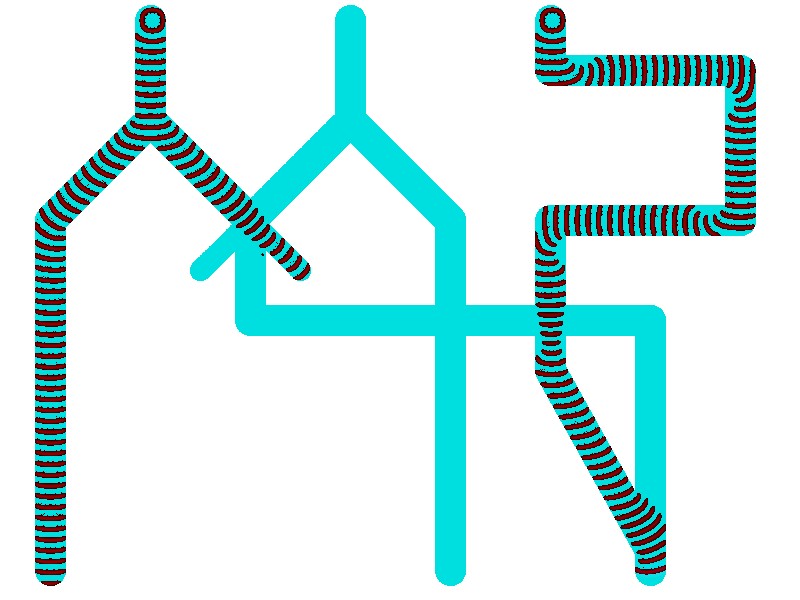}} 
\subfigure[]{\includegraphics[scale=0.2]{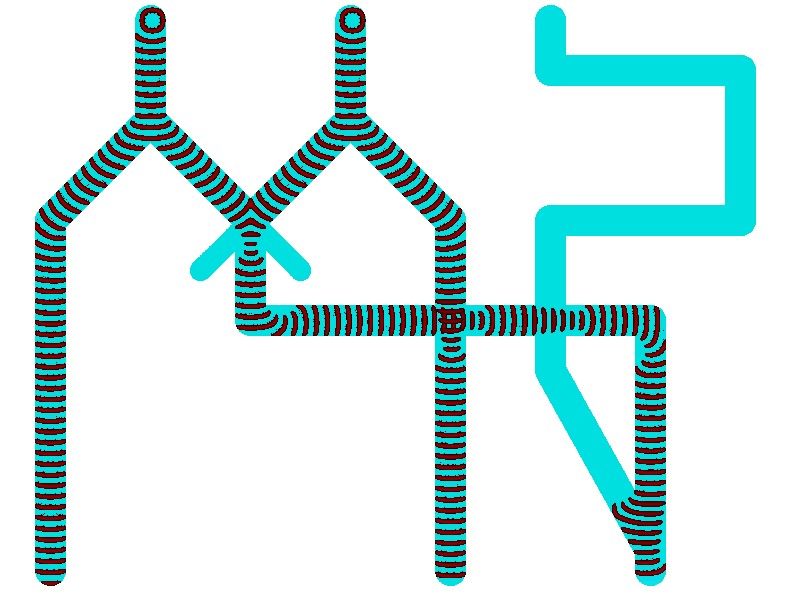}} 
\subfigure[]{\includegraphics[scale=0.2]{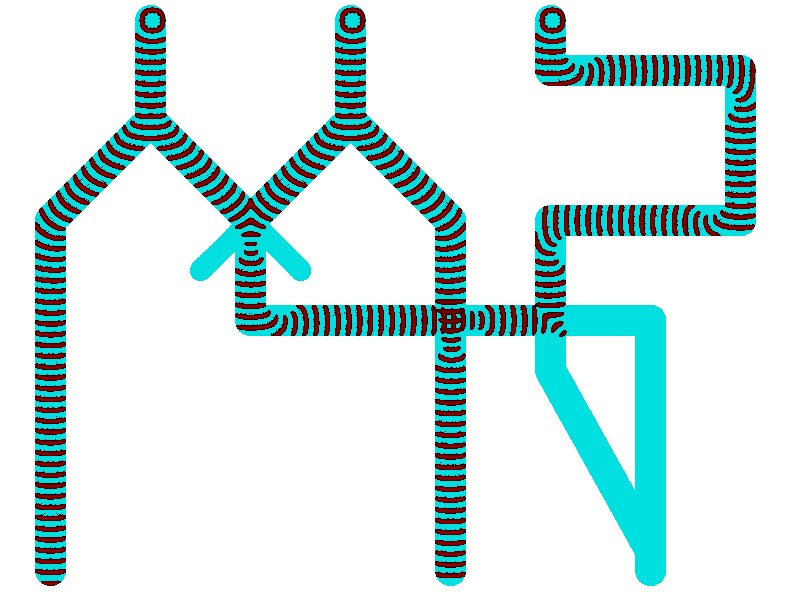}} 
\caption{Time lapsed snapshots of wave-fragments propagating in
  simulated BZ medium implementing  Toffoli gate. 
  (a)~$z=0, x=0, y=1$, 
  (b)~$z=0, x=1, y=0$,
  (c)~$z=0, x=1, y=1$,
  (d)~$z=1, x=0, y=0$,
  (e)~$z=1, x=0, y=1$,
  (f)~$z=1, x=1, y=0$,
  (g)~$z=1, x=1, y=1$.
    See videos at \protect\url{https://drive.google.com/open?id=0BzPSgPF_2eyUa2JwZ2FDTXBSdDQ}
}
\label{toffolisnapshots}
\end{figure*}

An excitable medium device implementing Toffoli gate (Fig.~\ref{toffoli}) consists of six excitable channels intersecting at six junctions 
four of which --- $j_3$, $j_4$, $j_6$ --- are sub-excitable.  

Input $z=0$, $x=0$, $y=1$. Excitation wave-fragment propagates from $y$ to $j_5$ (Fig.~\ref{toffolisnapshots}a). Junction $j_5$ is sub-excitable therefore the wave-fragment 
travels across the junction without spreading into lateral openings.  The excitation front reaches output $b$. 

Input $z=0$, $x=1$, $y=0$. Wave-front initiated in $x$ splits into two wave-front at excitable junction $j_2$. The wave travelling along $j_2 j_3$ enters the cul-de-sac and dies. 
The wave running along $j_2 j_4$ reaches the output $a$ (Fig.~\ref{toffolisnapshots}b). 

Input $z=0$, $x=1$, $y=1$.  Paths leading from $x$ to $a$ and from $y$ to $b$ do not  intersect, therefore wave-fragments originated in $x$ and $y$ do not interact.
Dynamics of excitation is an super-position of the excitation scenarios $z=0$, $x=0$, $y=1$ and   $z=0$, $x=1$, $y=0$ (Fig.~\ref{toffolisnapshots}c).

Input $z=1$, $x=0$, $y=0$. Excitation wave-front travels from $z$ to $j_1$ (Fig.~\ref{toffolisnapshots}d). At $j_1$ the wave-front splits into segment $j_1 c$, where it reaches output $c$, and segment $j_1 j_3$, where the wave-front dies in the cul-de-sac after crossing the junction. 

Input $z=1$, $x=0$, $y=1$. Wave-fronts originating in $z$ and $y$ do not interact.  The wave-front started in $z$ reaches output $c$, the wave-front initiated in $y$ reaches 
output $b$  (Fig.~\ref{toffolisnapshots}e).

Input $z=1$, $x=1$, $y=0$. Wave-fronts from $z$ and $x$ propagate to outputs $c$ and $a$, respectively. They also propagate along segments $j_1 j_3$ and $j_2 j_3$ and collide at 
junction $j_3$. They fuse and travel as a single wave-front along $j_3 j_4$ and $j_4 j_5$; this wave-front appears at output $b$ (Fig.~\ref{toffolisnapshots}f). Not that traces of wave-fragments along paths $j_3$ to $b$ and $j_2$ to $a$ intersect on the time-lapsed snapshot  (Fig.~\ref{toffolisnapshots}f) however wave-fragments do not collide at junction $j_4$ because they enter the junction at different times.

Input $z=1$, $x=1$, $y=1$. Wave-fronts from $z$ and $x$ propagate to outputs $c$ and $a$, respectively (Fig.~\ref{toffolisnapshots}g). They also propagate along segments $j_1 j_3$ and $j_2 j_3$ and collide at  junction $j_3$. They fuse and travel as a single wave-front along $j_3 j_4$ and $j_4 j_5$. Wave-front from $y$ travels towards $j_5$. Distance from $x$ (or $z$) to $j_5$ is the same distance from $y$ to $j_5$. Therefore wave-front travelling from $j_3$ to $j_5$ collides at junction $j_5$ with wave-front travelling from $y$ to $j_5$. These two wave-fronts merge into a single wave-front, which collides into a separation between segments $j_5 j_7$ and $j_5 j_6$, the wave-front annihilates (Fig.~\ref{toffolisnapshots}g).  
Not that traces of wave-fragments along paths $j_3$ to $b$ and $j_2$ to $a$ intersect on the time-lapsed snapshot  (Fig.~\ref{toffolisnapshots}g) however wave-fragments do not collide at junction $j_4$ because they enter the junction at different times.

Thus, we have $c=z$, $a=z$, $b=y \oplus (zx)$. Exact Boolean represented by a wave-fragment on each of the segments is shown in Tab.~\ref{segmentvalues}c.

\section{On physical reversibility}
\label{reversibility}

\begin{figure}[!tbp]
\centering
\subfigure[]{\includegraphics[scale=0.3]{MargolusGateScheme}}
\subfigure[]{\includegraphics[scale=0.3]{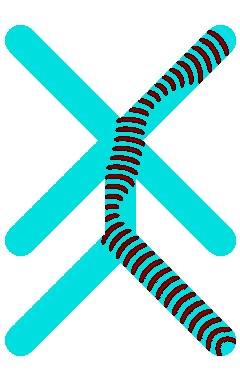}}
\subfigure[]{\includegraphics[scale=0.3]{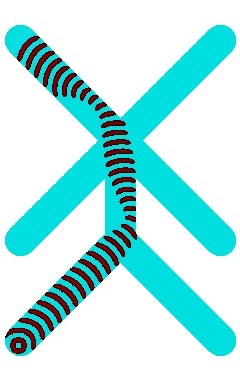}}
\subfigure[]{\includegraphics[scale=0.3]{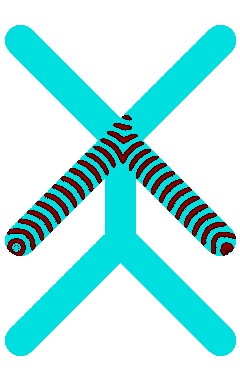}}
\subfigure[]{\includegraphics[scale=0.3]{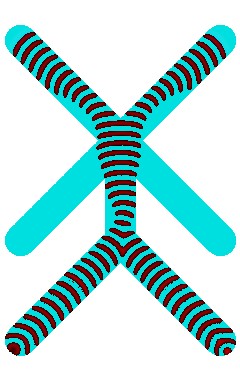}}
\caption{Time lapsed snapshots of wave-fragments propagating in
  simulated BZ medium implementing  reversed (inputs are swapped with outputs) 
  Margolus gate: $x=\overline{a}b+c$, $y=a\overline{b}+d$.
(a)~Scheme of the original gate. 
(b)~$a=0$, $b=0$, $c=0$, $d=1$,
(c)~$a=0$, $b=0$, $c=1$, $d=0$,
(d )~$a=1$, $b=1$, $c=0$, $d=0$,
(e)~$a=0$, $b=0$, $c=1$, $d=1$.
}
\label{margolusreversed}
\end{figure}

The designs of BZ implementation of collision-based gates are logically reversible but not physically reversible (under 'physical reversibility' we mean only that swapping inputs with outputs not reversibility of chemical reactions or excitation processes). Take for example implementation of Margolus gate (Fig.~\ref{margolus}ab).  Assume output channels in the 
original Margolus gate are inputs, and inputs in the original Margolus gate are outputs. If all inputs but $d$ are '0' 
excitation is initiated only in channel $d$ (Fig.~\ref{margolusreversed}b). The wave-fragment propagates into junction $j_2$ where it collides to a protruding non-excitable separator of 
segments $j_1 j_2$ and $j_2 c$. The wave-fragment changes its direction of movement in the result of collision and therefore propagates into segment 
$j_1 y$ (Fig.~\ref{margolusreversed}b).
Analogically, if only $c$ is excited the excitation wave-front exits the gate through output $y$  (Fig.~\ref{margolusreversed}c).  If $a$ is excited the wave-fragment propagates straight into output $y$; if $b$ is excited the wave-fragment propagates into $x$.  When both channels $a$ and $b$ are excited two excitation wave-fragments fuse into a single wave-fragments; this wave collides into a protruding separators between segments $j_1 x$ and $j_1 y$, and this wave annihilates (Fig.~\ref{margolusreversed}d). When $c$ and $d$ are excited the wave-front is formed in segment $j_2 j_1$. This wave-fragment splits into two wave-fragments propagating into segments $j_1 x$ and $j_1 y$. 
Thus, we have $x=\overline{a}b+c$ and $y=a\overline{b}+d$.

If we apply input signals to channels $a$, $b$, $c$ of Fredkin and Toffoli gates we get the following output signals on channels $x$, $y$, $z$. 

If inputs are swapped with output in the Fredkin gate (Fig.~\ref{fredkinscheme}) then $z=a+b+c$, $x=a$, $y=b$.  Wave-fronts representing signals generated at $a$, $b$ and $c$ 
do not interact with each other. If at least one of the channels $a$, $b$ or $c$ is excited the wave-front always propagates into channel $z$. Wave-front starting at $a$ splits at $j_4$ and goes into $x$ and $z$. Wave from $b$ splits  at $j_5$ and propagates to $z$ and $y$. Wave initiated at $c$ propagates along path $c$ to $j_2$ to $j_1$ to $z$.  
Thus we have $z=a+b+c$.  Excitation wave-front gets into $x$ only if channel $a$ is excited, thus $x=a$.  Channel $y$ gets excited only if excitation wave is initiated in channel $b$, thus $y=b$.

If inputs are swapped with outputs in Toffoli gate  (Fig.~\ref{toffoli}) then $z=b+c$, $x=a+b$, $y=b$.  Only excitation wave-front initiated at channel $b$ can excite $x$, $y$ and $z$. This is because excitation wave splits at the junction $j_6$ and thus excitation propagates along paths $b$ to $j_6$ to $j_7$ to $j_4$ (splits in $j_3$) to $z$ and $x$. At the same time, the wave propagates from $b$ to $j_6$ to $j_5$ to $y$. Channel $y$ is only excited from $b$, thus we have $y=b$. Channel $x$ can be excited by wave-fronts initiated at $a$ or at $b$, thus $x=a+b$. Channel $z$ is excited from $b$ or $c$, thus $z=b+c$.

\begin{figure*}[!tbp]
\centering
\subfigure[]{\includegraphics[scale=0.4]{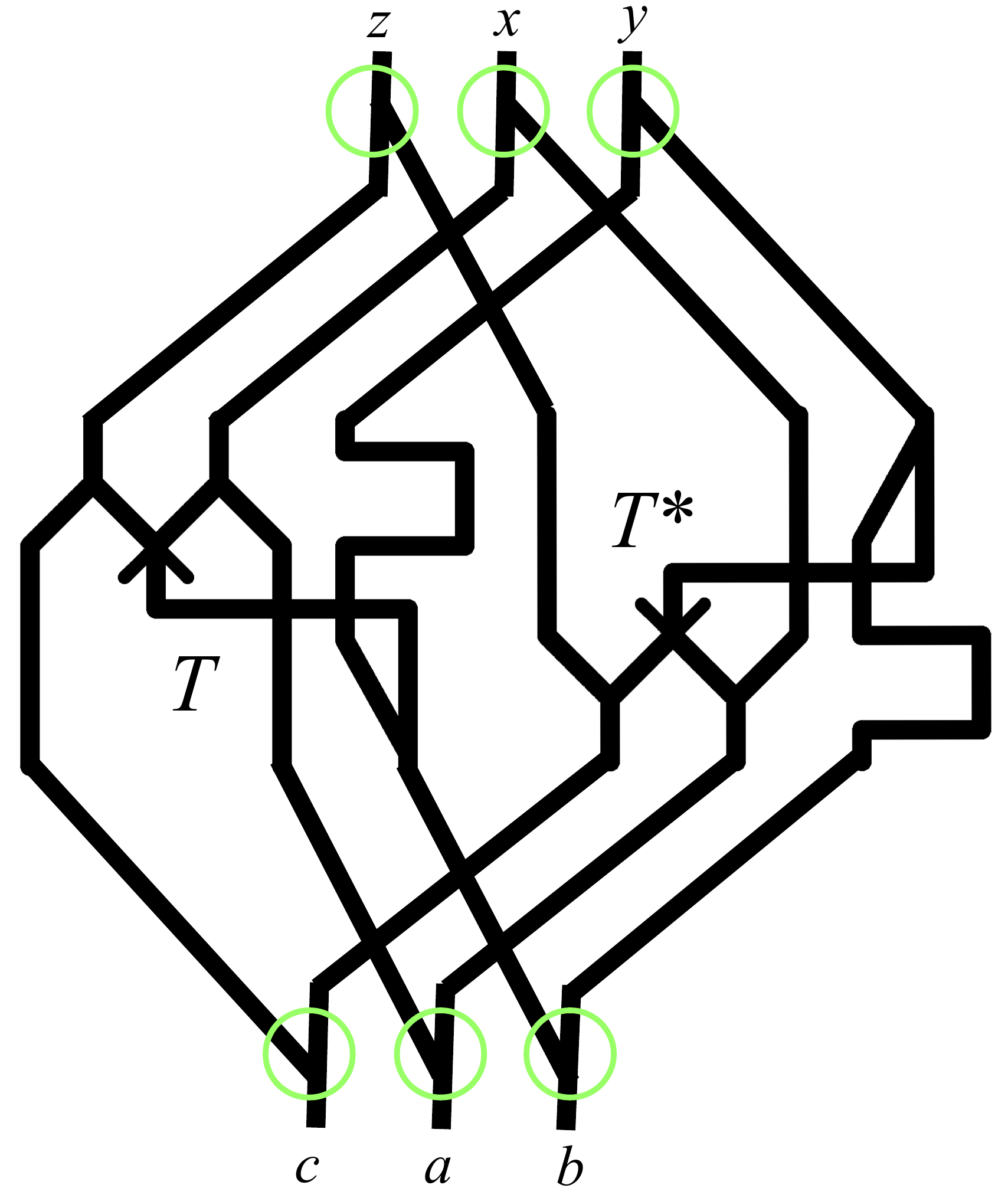}}\\
\subfigure[]{\includegraphics[scale=0.4]{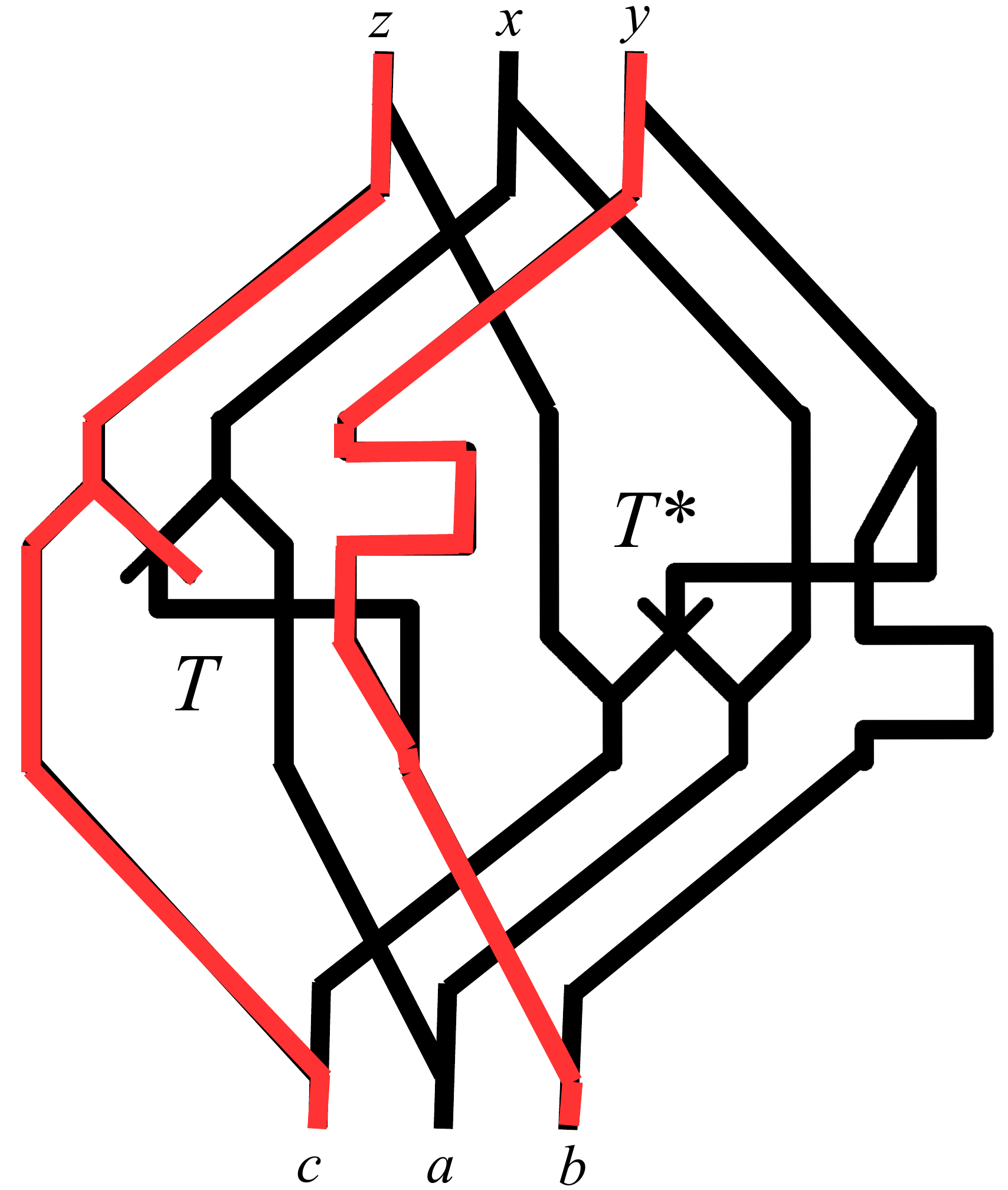}}
\subfigure[]{\includegraphics[scale=0.4]{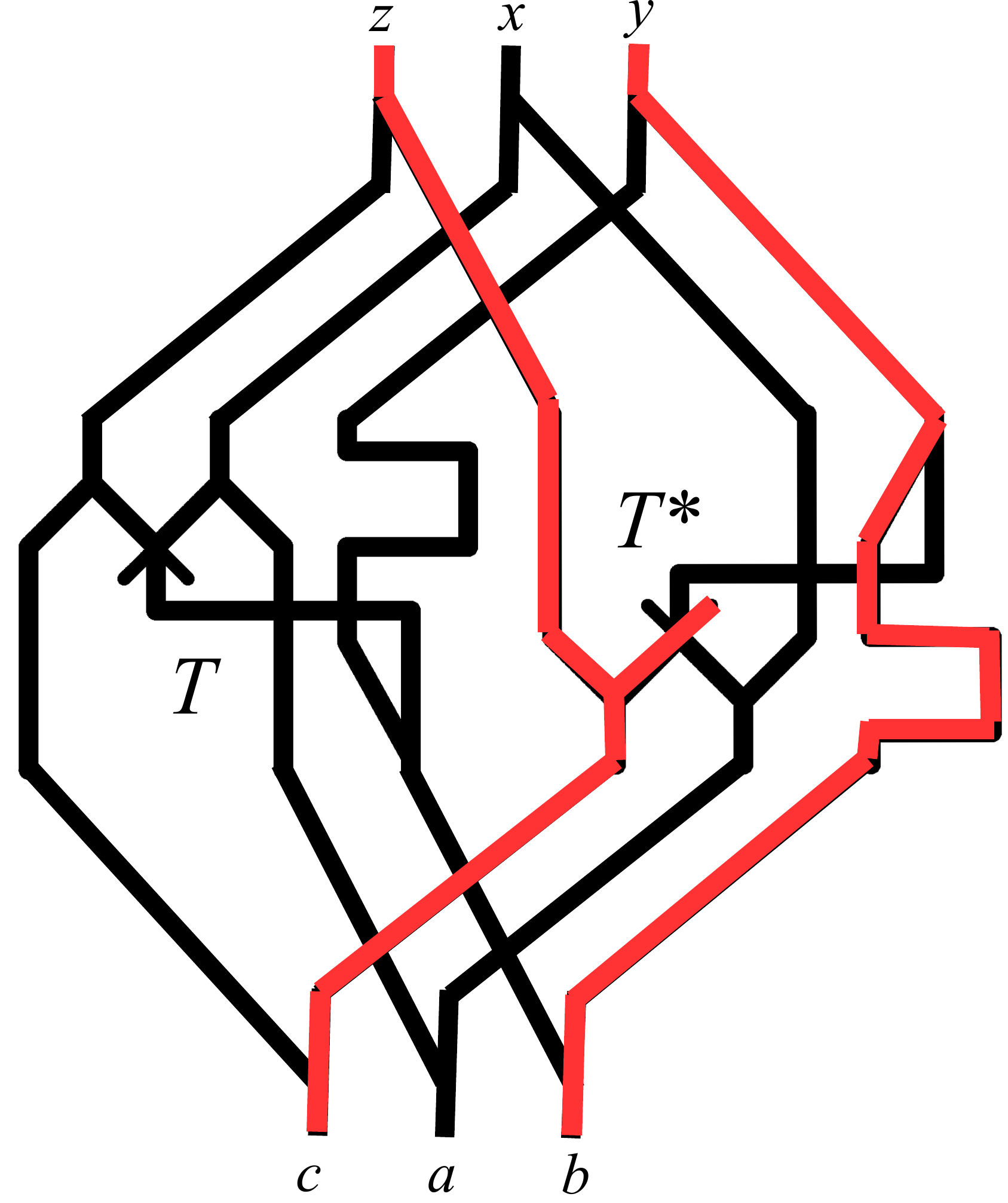}}
\caption{(a)~A scheme of physically reversible Toffoli gate made of two copies of original Toffoli gate Fig.~\ref{toffoli}.
Sub-excitable junctions, in addition to junctions shown in Fig.~\ref{toffoli}, are encircled. 
(b)~Trajectory of excitation wave-fronts for inputs $x=0$, $y=1$, $z=1$.
(c)~Trajectory of excitation wave-fronts for inputs $a=0$, $b=1$, $c=1$.}
\label{linkedgates}
\end{figure*}

\begin{figure}[!tbp]
\centering
\subfigure[]{\includegraphics[scale=0.4]{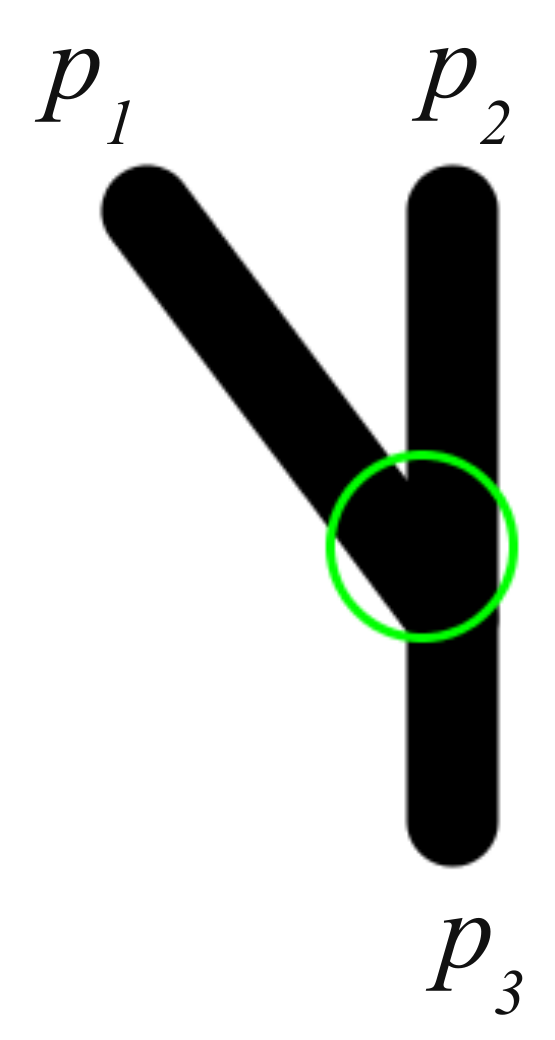}}
\subfigure[]{\includegraphics[scale=0.4]{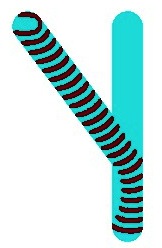}}
\subfigure[]{\includegraphics[scale=0.4]{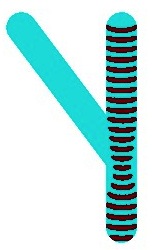}}
\subfigure[]{\includegraphics[scale=0.4]{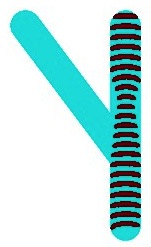}}
\subfigure[]{\includegraphics[scale=1]{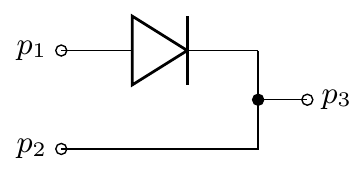}}
\caption{Junctions used to route input signals between sub-circuits in Fig.~\ref{linkedgates}. (a)~Scheme of a junction with ports $p_1$, $p_2$, $p_3$, domain of the junction inside circle
 is sub-excitable, other parts are excitable. (bcd)~Time-lapsed snapshots of excitation fronts: (b)~$p_1=1$, (c)~$p_2=1$, (d)~$p_3=1$, (e)~Equivalent electrical scheme.
See videos at \protect\url{https://drive.google.com/open?id=0BzPSgPF_2eyUb3lLSThwNVYzN3M}
}
\label{junctions}
\end{figure}

To make physically --- at a macro-level not at the level of chemical reactions --- reversible gates we can take two copies of the original  
gate (Fig.~\ref{toffoli}) flip one copy around an axis perpendicular to inputs and link these two sub-circuits in a single-circuit  
as shown in Fig.~\ref{linkedgates}.  Original copy of Toffoli gate is labelled $T$ in Fig.~\ref{linkedgates}, and flipped version $T^*$. 
A routing of input signals between $T$ and $T^*$ is implemented using sub-excitable junctions encircled in  Fig.~\ref{linkedgates}. 
The junction functioning is illustrated in Fig.~\ref{junctions}. When a wave-front is initiated either at port   $p_1$ (Fig.~\ref{junctions}b) or $p_2$ (Fig.~\ref{junctions}c) 
the wave-front propagates towards port $p_3$ without back-spreading into port $p_2$ ($p_1$). When a wave-front is initiated at port $p_3$ it propagates towards the port 
$p_2$ only (Fig.~\ref{junctions}d) but does not enter the channel leading to port $p_1$. In terms of electrical circuits, this is analogous of 
having  a diode at $p_1 p_3$ connection (Fig.~\ref{junctions}e).

Two examples of the circuit responding to $xyz$ and $abc$ inputs are shown in  Fig.~\ref{linkedgates}c. When signals enter the circuit via $a$, $b$ or $c$ the excitation waves propagate into the sub-circuit $T*$, the signals entered $x$, $y$ or $z$ are routed into the sub-circuit $T$. An example of traces of excitation propagation when signals are applied to inputs $x$, $y$, $z$ is shown in Fig.~\ref{linkedgates}b; and, when signals are applied to inputs $a$, $b$, $c$ in Fig.~\ref{linkedgates}c.

\section{Discussion}
\label{discussion}

The architectures of Fredkin and Toffoli gates proposed are evaluated in numerical model of a light sensitive Belousov-Zhabotinsky (BZ) reaction. Advantage of the light-sensitive BZ is that there is no need for altering a homogeneous substrate: an architecture of computing circuits is projected with light onto the medium such that conductive channels are non-illuminated and non-conductive domains are illuminated.   Further studies might also focus on mapping the designs  in molecular arrays~\cite{DBLP:journals/ijuc/Reif12, DBLP:journals/ijuc/KonkoliW14, valleau2012exciton, saikin2013photonics}; arrays of single-electron oscillators,, locally coupled with capacitors~\cite{oya2005reaction}, solid state reaction-diffusion devices with minority-carrier diffusion \cite{asai2001novel, takahashi2007cmos}; 
CNN chips, where pairs of layers represent activator and inhibitor concentrations, and diffusion and reaction are controlled via external bias 
voltages~\cite{shi2004spatial, arena1999realization, rekeczky2000stored, petras2003exploration, arena2004cnn, bonaiuto2000design, karahaliloglu2004mos}. 
Unconventional robotics is another application domain.  The BZ computing circuits can be integrated with self-propulsive BZ droplets~\cite{suematsu2016oscillation}. 
The BZ circuits can play a key role in  embedded parallel controllers for soft robots made of  
pH sensitive polymers and gels~\cite{yashin2006modeling, yoshida1999aspects, maeda2008control, yoshida2010self}, especially to implement decision-making circuits for 
crawling robots made of BZ gels~\cite{ren2016retrograde} as processors complimentary to  
a fluidic logic~\cite{rus2015design, nawaz2013unconventional, wehner2016integrated}.


\bibliographystyle{plain}
\bibliography{bibliography}

\begin{thebibliography}{100}

\bibitem{adamatzky2004collision}
Andrew Adamatzky.
\newblock Collision-based computing in {B}elousov--{Z}habotinsky medium.
\newblock {\em Chaos, Solitons \& Fractals}, 21(5):1259--1264, 2004.

\bibitem{adamatzky2010slime}
Andrew Adamatzky.
\newblock Slime mould logical gates: exploring ballistic approach.
\newblock {\em arXiv preprint arXiv:1005.2301}, 2010.

\bibitem{adamatzky2015binary}
Andrew Adamatzky.
\newblock Binary full adder, made of fusion gates, in a subexcitable
  {B}elousov-{Z}habotinsky system.
\newblock {\em Physical Review E}, 92(3):032811, 2015.

\bibitem{adamatzky2011polymorphic}
Andrew Adamatzky, Ben de~Lacy~Costello, and Larry Bull.
\newblock On polymorphic logical gates in subexcitable chemical medium.
\newblock {\em International Journal of Bifurcation and Chaos},
  21(07):1977--1986, 2011.

\bibitem{adamatzky2011towards}
Andrew Adamatzky, Ben De~Lacy~Costello, Larry Bull, and Julian Holley.
\newblock Towards arithmetic circuits in sub-excitable chemical media.
\newblock {\em Israel Journal of Chemistry}, 51(1):56--66, 2011.

\bibitem{adamatzky2002collision}
Andrew Adamatzky and Benjamin de~Lacy~Costello.
\newblock Collision-free path planning in the {B}elousov-{Z}habotinsky medium
  assisted by a cellular automaton.
\newblock {\em Naturwissenschaften}, 89(10):474--478, 2002.

\bibitem{adamatzky2007binary}
Andrew Adamatzky and Benjamin de~Lacy~Costello.
\newblock Binary collisions between wave-fragments in a sub-excitable
  {B}elousov--{Z}habotinsky medium.
\newblock {\em Chaos, Solitons \& Fractals}, 34(2):307--315, 2007.

\bibitem{adamatzky2004experimental}
Andrew Adamatzky, Benjamin de~Lacy~Costello, Chris Melhuish, and Norman
  Ratcliffe.
\newblock Experimental implementation of mobile robot taxis with onboard
  {B}elousov--{Z}habotinsky chemical medium.
\newblock {\em Materials Science and Engineering: C}, 24(4):541--548, 2004.

\bibitem{amemiya1998two}
Takashi Amemiya, Takao Ohmori, Masaru Nakaiwa, and Tomohiko Yamaguchi.
\newblock Two-parameter stochastic resonance in a model of the photosensitive
  {B}elousov-{Z}habotinsky reaction in a flow system.
\newblock {\em The Journal of Physical Chemistry A}, 102(24):4537--4542, 1998.

\bibitem{arena2004cnn}
Paolo Arena, Adriano Basile, Luigi Fortuna, and Mattia Frasca.
\newblock Cnn wave based computation for robot navigation planning.
\newblock In {\em Circuits and Systems, 2004. ISCAS'04. Proceedings of the 2004
  International Symposium on}, volume~5, pages V--500. IEEE, 2004.

\bibitem{arena1999realization}
Paolo Arena, Luigi Fortuna, and Marco Branciforte.
\newblock Realization of a reaction-diffusion cnn algorithm for locomotion
  control in an hexapode robot.
\newblock {\em Journal of VLSI signal processing systems for signal, image and
  video technology}, 23(2-3):267--280, 1999.

\bibitem{asai2001novel}
Tetsuya Asai, Yuusaku Nishimiya, and Yoshihito Amemiya.
\newblock A novel reaction-diffusion system based on minority-carrier transport
  in solid-state cmos devices.
\newblock In {\em Semiconductor Device Research Symposium, 2001 International},
  pages 141--144. IEEE, 2001.

\bibitem{azhand2014three}
Arash Azhand, Jan~Frederik Totz, and Harald Engel.
\newblock Three-dimensional autonomous pacemaker in the photosensitive
  {B}elousov-{Z}habotinsky medium.
\newblock {\em EPL (Europhysics Letters)}, 108(1):10004, 2014.

\bibitem{barenco1995elementary}
Adriano Barenco, Charles~H Bennett, Richard Cleve, David~P DiVincenzo, Norman
  Margolus, Peter Shor, Tycho Sleator, John~A Smolin, and Harald Weinfurter.
\newblock Elementary gates for quantum computation.
\newblock {\em Physical review A}, 52(5):3457, 1995.

\bibitem{beato2003pulse}
Valentina Beato and Harald Engel.
\newblock Pulse propagation in a model for the photosensitive
  {B}elousov-{Z}habotinsky reaction with external noise.
\newblock In {\em SPIE's First International Symposium on Fluctuations and
  Noise}, pages 353--362. International Society for Optics and Photonics, 2003.

\bibitem{belousov1959periodic}
Boris~P {B}elousov.
\newblock A periodic reaction and its mechanism.
\newblock {\em Compilation of Abstracts on Radiation Medicine}, 147(145):1,
  1959.

\bibitem{bennett1988notes}
Charles~H Bennett.
\newblock Notes on the history of reversible computation.
\newblock {\em {I}{B}{M}s Journal of Research and Development}, 32(1):16--23,
  1988.

\bibitem{berman1994quantum}
Gennady~P Berman, Gary~D Doolen, Darryl~D Holm, and Vladimir~I Tsifrinovich.
\newblock Quantum computer on a class of one-dimensional {I}sing systems.
\newblock {\em Physics Letters A}, 193(5-6):444--450, 1994.

\bibitem{bonaiuto2000design}
V~Bonaiuto, A~Maffucci, G~Miano, M~Salerno, F~Sargeni, and C~Visone.
\newblock Design of a cellular nonlinear network for analogue simulation of
  reaction-diffusion pdes.
\newblock In {\em Circuits and Systems, 2000. Proceedings. ISCAS 2000 Geneva.
  The 2000 IEEE International Symposium on}, volume~3, pages 431--434. IEEE,
  2000.

\bibitem{cerf1998optical}
NJ~Cerf, C~Adami, and PG~Kwiat.
\newblock Optical simulation of quantum logic.
\newblock {\em Physical Review A}, 57(3):R1477, 1998.

\bibitem{chang1982magnetic}
Hsu Chang.
\newblock Magnetic-bubble conservative logic.
\newblock {\em International Journal of Theoretical Physics}, 21(12):955--960,
  1982.

\bibitem{cory1998nuclear}
David~G Cory, Mark~D Price, and Timothy~F Havel.
\newblock Nuclear magnetic resonance spectroscopy: An experimentally accessible
  paradigm for quantum computing.
\newblock {\em Physica D: Nonlinear Phenomena}, 120(1):82--101, 1998.

\bibitem{costello2011towards}
Ben De~Lacy Costello, Andy Adamatzky, Ishrat Jahan, and Liang Zhang.
\newblock Towards constructing one-bit binary adder in excitable chemical
  medium.
\newblock {\em Chemical Physics}, 381(1):88--99, 2011.

\bibitem{cuykendall1987control}
Robert Cuykendall and Debra McMillin.
\newblock Control-specific optical {F}redkin circuits.
\newblock {\em Applied optics}, 26(10):1959--1963, 1987.

\bibitem{de2009implementation}
Ben de~Lacy~Costello, Rita Toth, Christopher Stone, Andrew Adamatzky, and Larry
  Bull.
\newblock Implementation of glider guns in the light-sensitive
  {B}elousov-{Z}habotinsky medium.
\newblock {\em Physical Review E}, 79(2):026114, 2009.

\bibitem{de2011reversible}
Alexis De~Vos.
\newblock {\em Reversible computing: fundamentals, quantum computing, and
  applications}.
\newblock John Wiley \& Sons, 2011.

\bibitem{dockery1988dispersion}
JD~Dockery, James~P Keener, and JJ~Tyson.
\newblock Dispersion of traveling waves in the {B}elousov-{Z}habotinskii
  reaction.
\newblock {\em Physica D: Nonlinear Phenomena}, 30(1):177--191, 1988.

\bibitem{escuela2014symbol}
Gabi Escuela, Gerd Gruenert, and Peter Dittrich.
\newblock Symbol representations and signal dynamics in evolving droplet
  computers.
\newblock {\em Natural Computing}, 13(2):247--256, 2014.

\bibitem{fei2002realization}
Xue Fei, Du~Jiang-Feng, Shi Ming-Jun, Zhou Xian-Yi, Han Rong-Dian, and
  Wu~Ji-Hui.
\newblock Realization of the {F}redkin gate by three transition pulses in a
  nuclear magnetic resonance quantum information processor.
\newblock {\em Chinese physics letters}, 19(8):1048, 2002.

\bibitem{field1974oscillations}
Richard~J Field and Richard~M Noyes.
\newblock Oscillations in chemical systems. iv. limit cycle behavior in a model
  of a real chemical reaction.
\newblock {\em The Journal of Chemical Physics}, 60(5):1877--1884, 1974.

\bibitem{fratto2015biomolecular}
Brian~E Fratto, Nataliia Guz, and Evgeny Katz.
\newblock Biomolecular computing realized in parallel flow systems:
  Enzyme-based double feynman logic gate.
\newblock {\em Parallel Processing Letters}, 25(01):1540001, 2015.

\bibitem{fratto2015reversible}
Brian~E Fratto and Evgeny Katz.
\newblock Reversible logic gates based on enzyme-biocatalyzed reactions and
  realized in flow cells: A modular approach.
\newblock {\em ChemPhysChem}, 16(7):1405--1415, 2015.

\bibitem{fratto2016controlled}
Brian~E Fratto and Evgeny Katz.
\newblock Controlled logic gates�switch gate and fredkin gate based on
  enzyme-biocatalyzed reactions realized in flow cells.
\newblock {\em ChemPhysChem}, 2016.

\bibitem{fratto2016utilization}
Brian~E Fratto and Evgeny Katz.
\newblock Utilization of a fluidic infrastructure for the realization of
  enzyme-based {B}oolean logic operations.
\newblock {\em International Journal of Parallel, Emergent and Distributed
  Systems}, pages 1--18, 2016.

\bibitem{fredkin2002conservative}
Edward {F}redkin and Tommaso {T}offoli.
\newblock Conservative logic.
\newblock In {\em Collision-based computing}, pages 47--81. Springer, 2002.

\bibitem{gentili2012belousov}
Pier~Luigi Gentili, Viktor Horvath, Vladimir~K Vanag, and Irving~R Epstein.
\newblock Belousov-{Z}habotinsky ``chemical neuron'' as a binary and fuzzy
  logic processor.
\newblock {\em IJUC}, 8(2):177--192, 2012.

\bibitem{gorecka2003t}
J~Gorecka and J~Gorecki.
\newblock T-shaped coincidence detector as a band filter of chemical signal
  frequency.
\newblock {\em Physical Review E}, 67(6):067203, 2003.

\bibitem{gorecki2015chemical}
J~Gorecki, K~Gizynski, J~Guzowski, JN~Gorecka, P~Garstecki, G~Gruenert, and
  P~Dittrich.
\newblock Chemical computing with reaction--diffusion processes.
\newblock {\em Phil. Trans. R. Soc. A}, 373(2046):20140219, 2015.

\bibitem{gorecki2014information}
J~Gorecki, JN~Gorecka, and Andrew Adamatzky.
\newblock Information coding with frequency of oscillations in
  {B}elousov-{Z}habotinsky encapsulated disks.
\newblock {\em Physical Review E}, 89(4):042910, 2014.

\bibitem{gorecki2003chemical}
J~Gorecki, K~Yoshikawa, and Y~Igarashi.
\newblock On chemical reactors that can count.
\newblock {\em The Journal of Physical Chemistry A}, 107(10):1664--1669, 2003.

\bibitem{gorecki2006information}
Jerzy Gorecki and Joanna~Natalia Gorecka.
\newblock Information processing with chemical excitations--from instant
  machines to an artificial chemical brain.
\newblock {\em International Journal of Unconventional Computing}, 2(4), 2006.

\bibitem{gorecki2009information}
Jerzy Gorecki, Joanna~Natalia Gorecka, and Yasuhiro Igarashi.
\newblock Information processing with structured excitable medium.
\newblock {\em Natural Computing}, 8(3):473--492, 2009.

\bibitem{gruenert2014understanding}
Gerd Gruenert, Konrad Gizynski, Gabi Escuela, Bashar Ibrahim, Jerzy Gorecki,
  and Peter Dittrich.
\newblock Understanding networks of computing chemical droplet neurons based on
  information flow.
\newblock {\em International journal of neural systems}, page 1450032, 2014.

\bibitem{gruenert2015understanding}
Gerd Gruenert, Konrad Gizynski, Gabi Escuela, Bashar Ibrahim, Jerzy Gorecki,
  and Peter Dittrich.
\newblock Understanding networks of computing chemical droplet neurons based on
  information flow.
\newblock {\em International journal of neural systems}, 25(07):1450032, 2015.

\bibitem{digitalcomparator}
Shan Guo, Ming-Zhu Sun, and Xin Han.
\newblock Digital comparator in excitable chemical media.
\newblock {\em International Journal Unconventional Computing}, 2015.

\bibitem{hardy2007optics}
James Hardy and Joseph Shamir.
\newblock Optics inspired logic architecture.
\newblock {\em Optics Express}, 15(1):150--165, 2007.

\bibitem{hastings2016oregonator}
Harold~M Hastings, Richard~Jeffrey Field, Sabrina~Godfrey Sobel, and David
  Guralnick.
\newblock Oregonator scaling motivated by showalter-noyes limit.
\newblock {\em The Journal of Physical Chemistry A}, 2016.

\bibitem{hsu1994effects}
TJ~Hsu, CY~Mou, and DJ~Lee.
\newblock Effects of macromixing on the oregonator model of the
  belousov�zhabotinsky reaction in a stirred reactor.
\newblock {\em Chemical engineering science}, 49(24):5291--5305, 1994.

\bibitem{DBLP:journals/ijuc/IgarashiG11}
Yasuhiro Igarashi and Jerzy Gorecki.
\newblock Chemical diodes built with controlled excitable media.
\newblock {\em {IJUC}}, 7(3):141--158, 2011.

\bibitem{igarashi2006chemical}
Yasuhiro Igarashi, Jerzy Gorecki, and Joanna~Natalia Gorecka.
\newblock Chemical information processing devices constructed using a nonlinear
  medium with controlled excitability.
\newblock In {\em Unconventional Computation}, pages 130--138. Springer, 2006.

\bibitem{jahnke1989chemical}
W~Jahnke, WE~Skaggs, and Arthur~T Winfree.
\newblock Chemical vortex dynamics in the {B}elousov-{Z}habotinskii reaction
  and in the two-variable oregonator model.
\newblock {\em The Journal of Physical Chemistry}, 93(2):740--749, 1989.

\bibitem{kaminaga2006reaction}
Akiko Kaminaga, Vladimir~K Vanag, and Irving~R Epstein.
\newblock A reaction--diffusion memory device.
\newblock {\em Angewandte Chemie International Edition}, 45(19):3087--3089,
  2006.

\bibitem{karahaliloglu2004mos}
Koray Karahaliloglu and Sina Balkir.
\newblock An mos cell circuit for compact implementation of reaction-diffusion
  models.
\newblock In {\em Neural Networks, 2004. Proceedings. 2004 IEEE International
  Joint Conference on}, volume~1. IEEE, 2004.

\bibitem{katz2017enzyme}
Evgeny Katz and Brian~E Fratto.
\newblock Enzyme-based reversible logic gates operated in flow cells.
\newblock In {\em Advances in Unconventional Computing}, pages 29--59.
  Springer, 2017.

\bibitem{klein1999biomolecular}
Joshua~P Klein, Thomas~H Leete, and Harvey Rubin.
\newblock A biomolecular implementation of logically reversible computation
  with minimal energy dissipation.
\newblock {\em Biosystems}, 52(1):15--23, 1999.

\bibitem{DBLP:journals/ijuc/KonkoliW14}
Zoran Konkoli and G{\"{o}}ran Wendin.
\newblock On information processing with networks of nano-scale switching
  elements.
\newblock {\em {IJUC}}, 10(5-6):405--428, 2014.

\bibitem{kostinski2009experimental}
Natalie Kostinski, Mable~P Fok, and Paul~R Prucnal.
\newblock Experimental demonstration of an all-optical fiber-based {F}redkin
  gate.
\newblock {\em Optics letters}, 34(18):2766--2768, 2009.

\bibitem{krug1990analysis}
Hans~Juergen Krug, Ludwig Pohlmann, and Lothar Kuhnert.
\newblock Analysis of the modified complete oregonator accounting for oxygen
  sensitivity and photosensitivity of {B}elousov-{Z}habotinskii systems.
\newblock {\em Journal of Physical Chemistry}, 94(12):4862--4866, 1990.

\bibitem{kuhnert1986new}
L~Kuhnert.
\newblock A new optical photochemical memory device in a light-sensitive
  chemical active medium.
\newblock 1986.

\bibitem{kuhnert1989image}
Lothar Kuhnert, KI~Agladze, and VI~Krinsky.
\newblock Image processing using light-sensitive chemical waves.
\newblock 1989.

\bibitem{leporati2004simulating}
Alberto Leporati, Claudio Zandron, and Giancarlo Mauri.
\newblock Simulating the {F}redkin gate with energy-based p systems.
\newblock {\em J. UCS}, 10(5):600--619, 2004.

\bibitem{lloyd1993potentially}
Seth Lloyd et~al.
\newblock A potentially realizable quantum computer.
\newblock {\em Science}, 261(5128):1569--1571, 1993.

\bibitem{maeda2008control}
Shingo Maeda, Yusuke Hara, Ryo Yoshida, and Shuji Hashimoto.
\newblock Control of the dynamic motion of a gel actuator driven by the
  {B}elousov-{Z}habotinsky reaction.
\newblock {\em Macromolecular Rapid Communications}, 29(5):401--405, 2008.

\bibitem{margolus2002universal}
Norman Margolus.
\newblock Universal cellular automata based on the collisions of soft spheres.
\newblock In Andrew Adamatzky, editor, {\em Collision-based computing}, pages
  107--134. Springer, 2002.

\bibitem{milburn1989quantum}
GJ~Milburn.
\newblock Quantum optical {F}redkin gate.
\newblock {\em Physical Review Letters}, 62(18):2124, 1989.

\bibitem{moseley2014enzyme}
Fiona Moseley, Jan Hal{\'a}mek, Friederike Kramer, Arshak Poghossian, Michael~J
  Sch{\"o}ning, and Evgeny Katz.
\newblock An enzyme-based reversible cnot logic gate realized in a flow system.
\newblock {\em Analyst}, 139(8):1839--1842, 2014.

\bibitem{murali2002quantum}
KVRM Murali, Neeraj Sinha, TS~Mahesh, Malcolm~H Levitt, KV~Ramanathan, and Anil
  Kumar.
\newblock Quantum-information processing by nuclear magnetic resonance:
  Experimental implementation of half-adder and subtractor operations using an
  oriented spin-7/2 system.
\newblock {\em Physical Review A}, 66(2):022313, 2002.

\bibitem{nawaz2013unconventional}
Ahmad~Ahsan Nawaz, Xiaole Mao, Zackary~S Stratton, and Tony~Jun Huang.
\newblock Unconventional microfluidics: expanding the discipline.
\newblock {\em Lab on a Chip}, 13(8):1457--1463, 2013.

\bibitem{orbach2014dnazyme}
Ron Orbach, Francoise Remacle, RD~Levine, and Itamar Willner.
\newblock Dnazyme-based 2: 1 and 4: 1 multiplexers and 1: 2 demultiplexer.
\newblock {\em Chemical Science}, 5(3):1074--1081, 2014.

\bibitem{oya2005reaction}
Takahide Oya, Tetsuya Asai, Takashi Fukui, and Yoshihito Amemiya.
\newblock Reaction-diffusion systems consisting of single-electron oscillators.
\newblock {\em International Journal of Unconventional Computing}, 1(2):179,
  2005.

\bibitem{petras2003exploration}
Istv{\'a}n Petr{\'a}s, Csaba Rekeczky, Tam{\'a}s Roska, Ricardo Carmona,
  Francisco Jim{\'e}nez-Garrido, and Angel Rodr{\'\i}guez-V{\'a}zquez.
\newblock Exploration of spatial-temporal dynamic phenomena in a 32$\times$
  32-cell stored program two-layer cnn universal machine chip prototype.
\newblock {\em Journal of Circuits, Systems, and Computers}, 12(06):691--710,
  2003.

\bibitem{poustie2000demonstration}
AJ~Poustie and KJ~Blow.
\newblock Demonstration of an all-optical {F}redkin gate.
\newblock {\em Optics Communications}, 174(1):317--320, 2000.

\bibitem{pullela2009temperature}
Srinivasa~R Pullela, Diego Cristancho, Peng He, Dawei Luo, Kenneth~R Hall, and
  Zhengdong Cheng.
\newblock Temperature dependence of the oregonator model for the
  {B}elousov-{Z}habotinsky reaction.
\newblock {\em Physical Chemistry Chemical Physics}, 11(21):4236--4243, 2009.

\bibitem{rambidi2001chemical}
NG~Rambidi and D~Yakovenchuk.
\newblock Chemical reaction-diffusion implementation of finding the shortest
  paths in a labyrinth.
\newblock {\em Physical Review E}, 63(2):026607, 2001.

\bibitem{DBLP:journals/ijuc/Reif12}
John~H. Reif.
\newblock Local parallel biomolecular computation.
\newblock {\em {IJUC}}, 8(5-6):459--507, 2012.

\bibitem{rekeczky2000stored}
Csaba Rekeczky, T~Serrano-Gotarredona, Tam{\'a}s Roska, and
  A~Rodriguez-Vazquez.
\newblock A stored program 2 nd order/3-layer complex cell cnn-um.
\newblock In {\em Cellular Neural Networks and Their Applications, 2000.(CNNA
  2000). Proceedings of the 2000 6th IEEE International Workshop on}, pages
  213--217. IEEE, 2000.

\bibitem{ren2016retrograde}
Lin Ren, Weibing She, Qingyu Gao, Changwei Pan, Chen Ji, and Irving~R Epstein.
\newblock Retrograde and direct wave locomotion in a photosensitive
  self-oscillating gel.
\newblock {\em Angewandte Chemie}, 2016.

\bibitem{roy2010novel}
Sukhdev Roy and Mohit Prasad.
\newblock Novel proposal for all-optical {F}redkin logic gate with
  bacteriorhodopsin-coated microcavity and its applications.
\newblock {\em Optical Engineering}, 49(6):065201--065201, 2010.

\bibitem{rus2015design}
Daniela Rus and Michael~T Tolley.
\newblock Design, fabrication and control of soft robots.
\newblock {\em Nature}, 521(7553):467--475, 2015.

\bibitem{saikin2013photonics}
Semion~K Saikin, Alexander Eisfeld, St{\'e}phanie Valleau, and Al{\'a}n
  Aspuru-Guzik.
\newblock Photonics meets excitonics: natural and artificial molecular
  aggregates.
\newblock {\em Nanophotonics}, 2(1):21--38, 2013.

\bibitem{schumann2015conventional}
Andrew Schumann.
\newblock Conventional and unconventional reversible logic gates on physarum
  polycephalum.
\newblock {\em International Journal of Parallel, Emergent and Distributed
  Systems}, pages 1--14, 2015.

\bibitem{seelig2006enzyme}
Georg Seelig, David Soloveichik, David~Yu Zhang, and Erik Winfree.
\newblock Enzyme-free nucleic acid logic circuits.
\newblock {\em science}, 314(5805):1585--1588, 2006.

\bibitem{vsevcikova1996dynamics}
Hana {\v{S}}ev?{\'\i}kov{\'a}, Igor Schreiber, and Milo{\v{s}} Marek.
\newblock Dynamics of oxidation {B}elousov-{Z}habotinsky waves in an electric
  field.
\newblock {\em The Journal of Physical Chemistry}, 100(49):19153--19164, 1996.

\bibitem{shamir1986optical}
Joseph Shamir, H~John Caulfield, William Micelli, and Robert~J Seymour.
\newblock Optical computing and the {F}redkin gates.
\newblock {\em Applied optics}, 25(10):1604--1607, 1986.

\bibitem{shi2004spatial}
Bertram~E Shi and Tao Luo.
\newblock Spatial pattern formation via reaction-diffusion dynamics in
  32$\times$ 32$\times$ 4 cnn chip.
\newblock {\em IEEE Transactions on Circuits and Systems I: Regular Papers},
  51(5):939--947, 2004.

\bibitem{sielewiesiuk2001logical}
Jakub Sielewiesiuk and Jerzy G{\'o}recki.
\newblock Logical functions of a cross junction of excitable chemical media.
\newblock {\em The Journal of Physical Chemistry A}, 105(35):8189--8195, 2001.

\bibitem{smolin1996five}
John~A Smolin and David~P DiVincenzo.
\newblock Five two-bit quantum gates are sufficient to implement the quantum
  {F}redkin gate.
\newblock {\em Physical Review A}, 53(4):2855, 1996.

\bibitem{steinbock1996chemical}
Oliver Steinbock, Petteri Kettunen, and Kenneth Showalter.
\newblock Chemical wave logic gates.
\newblock {\em The Journal of Physical Chemistry}, 100(49):18970--18975, 1996.

\bibitem{steinbock1995navigating}
Oliver Steinbock, {\'A}gota T{\'o}th, and Kenneth Showalter.
\newblock Navigating complex labyrinths: optimal paths from chemical waves.
\newblock {\em Science}, pages 868--868, 1995.

\bibitem{stevens2012time}
William~M Stevens, Andrew Adamatzky, Ishrat Jahan, and Ben de~Lacy~Costello.
\newblock Time-dependent wave selection for information processing in excitable
  media.
\newblock {\em Physical Review E}, 85(6):066129, 2012.

\bibitem{stovold2012simulating}
James Stovold and Simon O�Keefe.
\newblock Simulating neurons in reaction-diffusion chemistry.
\newblock In {\em International Conference on Information Processing in Cells
  and Tissues}, pages 143--149. Springer, 2012.

\bibitem{stovold2016reaction}
James Stovold and Simon O�Keefe.
\newblock Reaction--diffusion chemistry implementation of associative memory
  neural network.
\newblock {\em International Journal of Parallel, Emergent and Distributed
  Systems}, pages 1--21, 2016.

\bibitem{stovold2017associative}
James Stovold and Simon O�Keefe.
\newblock Associative memory in reaction-diffusion chemistry.
\newblock In {\em Advances in Unconventional Computing}, pages 141--166.
  Springer, 2017.

\bibitem{suematsu2016oscillation}
Nobuhiko~J Suematsu, Yoshihito Mori, Takashi Amemiya, and Satoshi Nakata.
\newblock Oscillation of speed of a self-propelled belousov--zhabotinsky
  droplet.
\newblock {\em The Journal of Physical Chemistry Letters}, 7(17):3424--3428,
  2016.

\bibitem{sun2013multi}
Ming-Zhu Sun and Xin Zhao.
\newblock Multi-bit binary decoder based on {B}elousov-{Z}habotinsky reaction.
\newblock {\em The Journal of chemical physics}, 138(11):114106, 2013.

\bibitem{suncrossover}
Ming-Zhu Sun and Xin Zhao.
\newblock Crossover structures for logical computations in excitable chemical
  medium.
\newblock {\em International Journal Unconventional Computing}, 2015.

\bibitem{taboada1994spiral}
JJ~Taboada, AP~Munuzuri, V~P{\'e}rez-Mu{\~n}uzuri, M~G{\'o}mez-Gesteira, and
  V~P{\'e}rez-Villar.
\newblock Spiral breakup induced by an electric current in a
  belousov--zhabotinsky medium.
\newblock {\em Chaos: An Interdisciplinary Journal of Nonlinear Science},
  4(3):519--524, 1994.

\bibitem{takahashi2007cmos}
Motoyoshi Takahashi, Tetsuya Asai, Tetsuya Hirose, and Yoshihito Amemiya.
\newblock A cmos reaction-diffusion device using minority-carrier diffusion in
  semiconductors.
\newblock {\em International Journal of Bifurcation and Chaos},
  17(05):1713--1719, 2007.

\bibitem{takigawa2011dendritic}
Hisako Takigawa-Imamura and Ikuko~N Motoike.
\newblock Dendritic gates for signal integration with excitability-dependent
  responsiveness.
\newblock {\em Neural Networks}, 24(10):1143--1152, 2011.

\bibitem{toffoli1980reversible}
Tommaso {T}offoli.
\newblock Reversible computing.
\newblock In {\em International Colloquium on Automata, Languages, and
  Programming}, pages 632--644. Springer, 1980.

\bibitem{tomita1979chaos}
Kazuhisa Tomita and Ichiro Tsuda.
\newblock Chaos in the {B}elousov-{Z}habotinsky reaction in a flow system.
\newblock {\em Physics Letters A}, 71(5):489--492, 1979.

\bibitem{toth2009experimental}
Rita Toth, Christopher Stone, Andrew Adamatzky, Ben de~Lacy~Costello, and Larry
  Bull.
\newblock Experimental validation of binary collisions between wave fragments
  in the photosensitive {B}elousov--{Z}habotinsky reaction.
\newblock {\em Chaos, Solitons \& Fractals}, 41(4):1605--1615, 2009.

\bibitem{toth2010simple}
Rita Toth, Christopher Stone, Ben de~Lacy~Costello, Andrew Adamatzky, and Larry
  Bull.
\newblock Simple collision-based chemical logic gates with adaptive computing.
\newblock {\em Theoretical and Technological Advancements in Nanotechnology and
  Molecular Computation: Interdisciplinary Gains: Interdisciplinary Gains},
  page 162, 2010.

\bibitem{valleau2012exciton}
St{\'e}phanie Valleau, Semion~K Saikin, Man-Hong Yung, and Al{\'a}n~Aspuru
  Guzik.
\newblock Exciton transport in thin-film cyanine dye j-aggregates.
\newblock {\em The journal of chemical physics}, 137(3):034109, 2012.

\bibitem{vanag2000pattern}
Vladimir~K Vanag, Anatol~M Zhabotinsky, and Irving~R Epstein.
\newblock Pattern formation in the {B}elousov-{Z}habotinsky reaction with
  photochemical global feedback.
\newblock {\em The Journal of Physical Chemistry A}, 104(49):11566--11577,
  2000.

\bibitem{vavilin1968effect}
V.~A. Vavilin, A.~M. Zhabotinsky, and AN~Zaikin.
\newblock Effect of ultraviolet radiation on oscillating oxidation reaction of
  malonic acid derivatives, 1968.

\bibitem{DBLP:journals/ijuc/Vazquez-OteroFDD14}
Alejandro Vazquez{-}Otero, Jan Faigl, Natividad Duro, and Raquel Dormido.
\newblock Reaction-diffusion based computational model for autonomous mobile
  robot exploration of unknown environments.
\newblock {\em {IJUC}}, 10(4):295--316, 2014.

\bibitem{wehner2016integrated}
Michael Wehner, Ryan~L Truby, Daniel~J Fitzgerald, Bobak Mosadegh, George~M
  Whitesides, Jennifer~A Lewis, and Robert~J Wood.
\newblock An integrated design and fabrication strategy for entirely soft,
  autonomous robots.
\newblock {\em Nature}, 536(7617):451--455, 2016.

\bibitem{wei2013compact}
Hai-Rui Wei and Fu-Guo Deng.
\newblock Compact quantum gates on electron-spin qubits assisted by diamond
  nitrogen-vacancy centers inside cavities.
\newblock {\em Physical Review A}, 88(4):042323, 2013.

\bibitem{wenzler2013nanomechanical}
Josef-Stefan Wenzler, Tyler Dunn, Tommaso {T}offoli, and Pritiraj Mohanty.
\newblock A nanomechanical {F}redkin gate.
\newblock {\em Nano letters}, 14(1):89--93, 2013.

\bibitem{winfree1989three}
Arthur~T Winfree and Wolfgang Jahnke.
\newblock Three-dimensional scroll ring dynamics in the
  {B}elousov-{Z}habotinskii reagent and in the two-variable oregonator model.
\newblock {\em The Journal of Physical Chemistry}, 93(7):2823--2832, 1989.

\bibitem{wood2004fredkin}
David~Harlan Wood and Chen Junghuei.
\newblock Fredkin gate circuits via recombination enzymes.
\newblock In {\em Congress on evolutionary computation}, 2004.

\bibitem{yashin2006modeling}
Victor~V Yashin and Anna~C Balazs.
\newblock Modeling polymer gels exhibiting self-oscillations due to the
  {B}elousov-{Z}habotinsky reaction.
\newblock {\em Macromolecules}, 39(6):2024--2026, 2006.

\bibitem{yokoi2004excitable}
Hiroshi Yokoi, Andy Adamatzky, Ben de~Lacy~Costello, and Chris Melhuish.
\newblock Excitable chemical medium controller for a robotic hand: Closed-loop
  experiments.
\newblock {\em International Journal of Bifurcation and Chaos},
  14(09):3347--3354, 2004.

\bibitem{yoshida2010self}
Ryo Yoshida.
\newblock Self-oscillating gels driven by the {B}elousov--{Z}habotinsky
  reaction as novel smart materials.
\newblock {\em Advanced Materials}, 22(31):3463--3483, 2010.

\bibitem{yoshida1999aspects}
Ryo Yoshida, Satoko Onodera, Tomohiko Yamaguchi, and Etsuo Kokufuta.
\newblock Aspects of the {B}elousov-{Z}habotinsky reaction in polymer gels.
\newblock {\em The Journal of Physical Chemistry A}, 103(43):8573--8578, 1999.

\bibitem{DBLP:journals/ijuc/YoshikawaMIYIGG09}
Kenichi Yoshikawa, Ikuko Motoike, T.~Ichino, T.~Yamaguchi, Yasuhiro Igarashi,
  Jerzy Gorecki, and Joanna~Natalia Gorecka.
\newblock Basic information processing operations with pulses of excitation in
  a reaction-diffusion system.
\newblock {\em {IJUC}}, 5(1):3--37, 2009.

\bibitem{zardalidis2004design}
George~T Zardalidis and Ioannis Karafyllidis.
\newblock Design and simulation of a nanoelectronic single-electron universal
  {F}redkin gate.
\newblock {\em IEEE Transactions on Circuits and Systems I: Regular Papers},
  51(12):2395--2403, 2004.

\bibitem{zhabotinsky1964periodic}
AM~{Z}habotinsky.
\newblock Periodic processes of malonic acid oxidation in a liquid phase.
\newblock {\em Biofizika}, 9(306-311):11, 1964.

\bibitem{zhang2012towards}
Guo-Mao Zhang, Ieong Wong, Meng-Ta Chou, and Xin Zhao.
\newblock Towards constructing multi-bit binary adder based on
  {B}elousov-{Z}habotinsky reaction.
\newblock {\em The Journal of chemical physics}, 136(16):164108, 2012.

\bibitem{zheng2013implementation}
Shi-Biao Zheng.
\newblock Implementation of {T}offoli gates with a single asymmetric
  {H}eisenberg x y interaction.
\newblock {\em Physical Review A}, 87(4):042318, 2013.

\end{thebibliography}

\end{document}